\documentclass[conference]{IEEEtran}
\IEEEoverridecommandlockouts
\usepackage{enumitem}
\usepackage[ruled,vlined,linesnumbered]{algorithm2e}
\usepackage{tabularx,booktabs}
\usepackage{amsmath,amssymb,amsfonts}
\usepackage{algorithmic}
\usepackage{graphicx}
\usepackage{textcomp}
\usepackage{xcolor}
\def\BibTeX{{\rm B\kern-.05em{\sc i\kern-.025em b}\kern-.08em
    T\kern-.1667em\lower.7ex\hbox{E}\kern-.125emX}}
\usepackage{biblatex}
\begin{document}

\title{DistrEdge: Speeding up Convolutional Neural Network Inference on Distributed Edge Devices}

\author{
    \IEEEauthorblockN{Xueyu Hou\IEEEauthorrefmark{1}, Yongjie Guan\IEEEauthorrefmark{1}, Tao Han\IEEEauthorrefmark{1}, Ning Zhang\IEEEauthorrefmark{2}}
    \IEEEauthorblockA{\IEEEauthorrefmark{1} New Jersey Institute of Technology, USA, \IEEEauthorrefmark{2} Windsor University, Canada}
    \{xh29, yg274, tao.han\}@njit.edu, \{ning.zhang\}@uwindsor.ca
}

\maketitle

\begin{abstract}
As the number of edge devices with computing resources (e.g., embedded GPUs, mobile phones, and laptops) increases, recent studies demonstrate that it can be beneficial to collaboratively run convolutional neural network (CNN) inference on more than one edge device. However, these studies make strong assumptions on the devices' conditions, and their application is far from practical. In this work, we propose a general method, called \textit{DistrEdge}, to provide CNN inference distribution strategies in environments with multiple IoT edge devices. By addressing heterogeneity in devices, network conditions, and nonlinear characters of CNN computation, DistrEdge is adaptive to a wide range of cases (e.g., with different network conditions, various device types) using deep reinforcement learning technology. We utilize the latest embedded AI computing devices (e.g., NVIDIA Jetson products) to construct cases of heterogeneous devices' types in the experiment. Based on our evaluations, DistrEdge can properly adjust the distribution strategy according to the devices' computing characters and the network conditions. It achieves 1.1 to 3$\times$ speedup compared to state-of-the-art methods.
\end{abstract}


\section{Introduction}
With the development of deep learning, various applications are utilizing deep neural networks (DNNs). Convolutional neural networks (CNNs) are one of the most popular types of DNNs, and are applied to a wide range of computer vision tasks like image classification~\cite{simonyan2014very, he2016deep, szegedy2016rethinking}, object detection~\cite{ren2015faster, redmon2016yolo9000}, and pose detection~\cite{cao2017realtime}. 
As CNN models have high computational complexity, it is challenging to run CNN model inference for edge devices~\cite{yao2018deep} like mobile phones, surveillance cameras, drones. Studies on CNN compression~\cite{zhang2015accelerating, han2015learning, he2018amc} reduce the computational complexity of a CNN model by sacrificing accuracy. Another popular solution is to offload the whole or partial CNN model to a server~\cite{kang2017neurosurgeon, jeong2018ionn, teerapittayanon2017distributed, yao2020deep, zhang2021dynamic}. However, their dependency on servers and remote network transmission leads to problems like trustworthiness, security, and privacy~\cite{wu2019enabling, enkhtaivan2020mediating, nasution2020iot}. 

Distributing CNN inference on multiple edge devices in a paralleling manner has been studied in~\cite{zeng2020coedge, mao2017modnn, mao2017mednn, zhao2018deepthings, stahl2021deeperthings, zhou2019adaptive, hadidi2018distributed, hadidi2019collaborative, hadidi2020toward, de2018partitioning, martins2019partitioning, de2015dianne} and demonstrate its benefits in reducing inference latency, avoiding overflow of memory/power on devices, and utilizing idle computing resources. Different from distributed CNN training on servers, CNN inference distribution on edge devices shows the following characters (detailed discussion in Section II): (1) much lower throughput between devices; (2) heterogeneous computing devices; (3) heterogeneous network throughput conditions. However, as will be discussed in Section II, state-of-the-art studies on CNN inference distribution~\cite{zeng2020coedge, mao2017modnn, mao2017mednn, zhao2018deepthings, stahl2021deeperthings, zhou2019adaptive, hadidi2018distributed, hadidi2019collaborative, hadidi2020toward, de2018partitioning, martins2019partitioning, de2015dianne} address these characters with strong assumptions and significantly limit their applicable cases in practice. Our method (DistrEdge) addresses these characters with Deep Reinforcement Learning (DRL) and reaches $1.1$ to $3 \times$ speedup compared to state-of-the-art methods.

The CNN inference distribution method we propose is called \textit{DistrEdge}. Given a group of edge devices and a CNN model for inference, DistrEdge can provide a distribution solution to run inference on these devices by comprehensively analyzing the CNN layer configurations, network conditions, and devices' computing characters. Compared to state-of-the-art studies, we address two practical facts. First, the edge devices available for CNN inference distribution can be heterogeneous in devices' types and network conditions. Second, the relationship between computing latency and CNN layer configurations can be highly nonlinear on edge devices~\cite{yao2018fastdeepiot}, referred to as \textit{nonlinear} character in this paper. One note is that the memory constraint is not considered in this paper because the latest embedded AI computing devices (e.g., NVIDIA Jetson) has adequate memory~\cite{jetson_products, bianco2018benchmark} and the memory consumption by CNN inference does not cause severe issues on them. Overall, the contributions of this paper are as follows:


\begin{itemize}[leftmargin=*]
  \item As far as we know, DistrEdge is the first CNN inference distribution method that can apply to distribution on heterogeneous edge devices with \textit{nonlinear} characters.
  \item As far as we know, DistrEdge is the first CNN inference distribution method that models the \textit{split} process as a Markov Decision Process (MDP) and utilizes Deep Reinforcement Learning (DRL) to make optimal split decisions.
  \item We evaluate the performance of CNN inference on edge devices, including the latest AI computing devices of NVIDIA Jetson products~\cite{jetson_products}. In comparison, the experiments in state-of-the-art studies limit their test cases with low-speed edge devices like Raspberry Pi3/4 (e.g.,~\cite{zhao2018deepthings, hadidi2019collaborative, zhou2019adaptive}) or Nexus 5 (e.g.,~\cite{mao2017modnn,mao2017mednn}).
  \item In the experiment, DistrEdge can achieve $1.1 \times$ to $3 \times$ speedup over state-of-the-art CNN inference distribution methods. 
\end{itemize}

\section{Related Work}
\subsection{Model Parallelism in DNN Training}
State-of-the-art model parallelism studies mainly focus on accelerating the training of DNNs on distributed server devices. In general, there are two types of model parallelism in training DNNs. The first type is graph partitioning~\cite{blelloch1999pipelining, giacomoni2008fastforward, gordon2006exploiting, gaunt2017ampnet, guan2019xpipe, harlap2018pipedream, chen2018efficient, narayanan2021memory, park2020hetpipe, tanaka2021automatic}. Due to the sequential dependency between the computation tasks in a deep neural network, most graph partitioning works use pipeline parallelism~\cite{giacomoni2008fastforward, gordon2006exploiting, tanaka2021automatic}. By splitting training batch into micro-batches, the sequential computation of each micro-batch can be pipelined with those of the other micro-batches~\cite{tanaka2021automatic}. The second type is tensor splitting~\cite{shazeer2018mesh, shoeybi2019megatron, jia2018exploring, jia2018beyond, wang2019supporting}. The tensor splitting adjusts the input/output data placement on the devices and splits corresponding computation.

Though the model parallelism has been a popular way in implementing DNN training on multiple devices in a distributed manner~\cite{gaunt2017ampnet, guan2019xpipe, harlap2018pipedream, chen2018efficient, narayanan2021memory, park2020hetpipe, tanaka2021automatic}, it is important to note that the devices on which the training is distributed usually refer to computing units like GPUs connected to one server cluster through high-bandwidth connections like NVLinks and InfiniBand~\cite{narayanan2021memory, park2020hetpipe,tanaka2021automatic}. These computing units are controlled by a single operating system on the cluster. The communication throughput between these computing units is $>$10Gbps~\cite{narayanan2021memory, park2020hetpipe,tanaka2021automatic}. In contrast, distributed edge devices refer to embedded computing boards like NVIDIA Jetson and Raspberry Pi connected through a wireless network. Each computing board has its own operating system, and the data transmission between these boards is through network protocols like TCP/IP. The communication throughput between these boards is usually $1Mbps$ to $1Gbps$, which is much lower than that between computing units on a cluster. Moreover, there is additional I/O reading and writing delay when transmitting data from one edge device to another. 
In addition, edge devices are usually responsible for DNN inference instead of training. One feature of DNN inference is that the input data usually with a batch size of $1$, and inference requests arrive randomly or periodically. Therefore, pipeline parallelism of training batches is not applicable in the distribution of DNN inference on edge devices.

\subsection{CNN Inference on Distributed Edge Devices}

Distributed CNN inference studies adopt tensor splitting in model parallelism. By splitting the input/output data (and corresponding computation) of layers in a model, these works demonstrate its benefits in reducing inference latency, avoiding overflow of memory/power on devices, and utilizing idle computing resources. However, state-of-the-art studies limit their applicable cases.~\cite{de2018partitioning, martins2019partitioning, de2015dianne} limit their implementation on a tiny CNN model like LeNet as they make distribution decisions for each neuron (most popular CNN models consist of a vast amount of neurons).~\cite{hadidi2018distributed, hadidi2019collaborative, hadidi2020toward} aim at splitting a specific layer in a CNN model equally and allocating the split parts on homogeneous edge devices to avoid memory overflow. For works that can be applied to the distribution of popular CNN models: ~\cite{zeng2020coedge, mao2017modnn, mao2017mednn} distribute CNN inference in a layer-by-layer way and split each layer in a CNN model based on the edge devices' computing capability~\cite{zeng2020coedge, mao2017modnn, mao2017mednn} and the network throughput~\cite{zeng2020coedge};~\cite{zhao2018deepthings, stahl2021deeperthings, zhou2019adaptive} fuse multiple layers together and split the fused layers to reduce the frequency of data transmission between distributed edge devices. Though fusing layers reduces transmission latency through a wireless network and I/O reading/wring latency,~\cite{zhao2018deepthings, stahl2021deeperthings, zhou2019adaptive} limit their applicable cases as homogeneous edge devices~\cite{zhao2018deepthings, stahl2021deeperthings} or devices with linear characters~\cite{zhou2019adaptive}. 

Overall, state-of-the-art studies implement CNN inference distribution with strong assumptions: (1) DeepThings~\cite{zhao2018deepthings} and DeeperThings~\cite{stahl2021deeperthings} assume that the edge devices for distribution are homogeneous and the layer(s) are split equally; (2) CoEdge~\cite{zeng2020coedge}, MoDNN~\cite{mao2017modnn}, MeDNN~\cite{mao2017mednn}, and AOFL~\cite{zhou2019adaptive} assume that the relationship between the computing latency and the layer configuration is linear and their ratio can be represented by a value (i.e., computing capability); (3) DeepThings~\cite{zhao2018deepthings}, DeeperThings~\cite{stahl2021deeperthings}, MoDNN~\cite{mao2017modnn}, and MeDNN~\cite{mao2017mednn} do not consider the network conditions of edge devices when making distribution decisions; (4) CoEdge~\cite{zeng2020coedge} and AOFL~\cite{zhou2019adaptive} assume the transmission latency to be proportional to the value of network throughput and the layer configuration, and calculate the split ratio based on the linear models of the computing and network transmission latency. However, as demonstrated in~\cite{yao2018fastdeepiot}, the relationship between computing latency and layer configurations can be highly nonlinear. In addition, the delay caused by I/O reading/writing should also be accounted for in the transmission latency, and calculating the transmission latency purely by the network throughput can be inaccurate. 

This paper develops a novel distribution method, DistrEdge, to break the above limitations of state-of-the-art works. Overall, DistrEdge reaches $1.1$ to $3 \times$ speedup compared to state-of-the-art distribution solutions in heterogeneous edge devices and network conditions without assuming a linear relationship between the computing/transmission latency and the layer configuration. One note is that, for CNN inference on distributed edge devices, we refer to the study that is distributing existing CNN models \textit{without} any modification of the models' architecture. The benefit of it is that it does not require additional training to recover the accuracy because the accuracy performance is \textit{not} affected in the distribution process. 

\subsection{Resource Allocation for CNN Inference}
Another related topic for CNN inference on distributed edge devices is to study the resource allocation for CNN inference tasks~\cite{mohammed2020distributed, chen2019iraf, sahni2018data}, they either regard one inference process as a black-box~\cite{chen2019iraf, sahni2018data} or split a CNN layer in the same way~\cite{mohammed2020distributed} as in~\cite{mao2017modnn}. RL-PDNN~\cite{baccour2021rl} targets improving the overall performance of DNN requests from multiple users, and it does not study model parallelism. As they focus on resource allocation for multiple DNN inference tasks, they are out of the scope of this paper's focus. 

\subsection{Architecture Search/Modification for Distributed CNN}
Different from the studies of CNN inference on distributed edge devices,~\cite{kim2017splitnet, bhardwaj2019memory, hadidi2020reducing, hadidi2020lcp, zhang2020adaptive} modify/search for neural architectures that are suitable for distributing on edge devices. The main focus of such studies is to eliminate data dependency between edge devices during the inference of CNN models.~\cite{kim2017splitnet, bhardwaj2019memory, hadidi2020lcp, zhang2020adaptive} propose different methods to modify existing CNN models based on: hierarchical grouping of categories for image classification~\cite{kim2017splitnet}; training several disjoint student models using knowledge distillation~\cite{bhardwaj2019memory}; tailoring intermediate connections between neurons to form a low-communication-parallelization version of a model~\cite{hadidi2020lcp}; decomposing the spatial dimension by clipping and quantization~\cite{zhang2020adaptive}.~\cite{hadidi2020reducing} proposes to search for novel neural architecture to provide higher concurrency and distribution opportunities. Though modifying/searching neural architecture for CNN distribution can achieve highly efficient parallel neural architecture, it intrinsically shows the following drawbacks: (1) it takes a long time to modify/search optimal neural architecture and to train the neural network in order to recover the accuracy; (2) The modified neural network cannot achieve the same level of accuracy even after retraining~\cite{kim2017splitnet, bhardwaj2019memory, hadidi2020reducing, hadidi2020lcp, zhang2020adaptive}. 

\subsection{Deep Reinforcement Learning in CNN Model Compression}
To tackle the complex configurations and performance of CNN architecture, DRL has been used in CNN model compression and is proven to be effective in finding optimal CNN model compression schemes~\cite{he2018amc, ashok2017n2n, zhan2021deep, soni2019hmc, yu2021gnn}. The configurations of the current layer~\cite{he2018amc, ashok2017n2n, zhan2021deep} and the action of the previous layer~\cite{he2018amc} are defined as the state in the DRL. The action is the compression decision on the current layer~\cite{he2018amc, ashok2017n2n, zhan2021deep, soni2019hmc, yu2021gnn}. The reward is defined as a function of the performance (e.g., accuracy)~\cite{he2018amc, ashok2017n2n, zhan2021deep, soni2019hmc, yu2021gnn}. As far as we know, DistrEdge is the first paper that utilizes DRL to find a CNN inference distribution strategy.

\section{Definitions and Challenges}


\subsection{Terms Definition}

For the convenience of the following discussion, we define the following terms: (1) \textit{layer-volume}: a layer-volume refers to one or more sequentially-connected layers in a CNN model (the same concept as fused layers in~\cite{zhao2018deepthings, stahl2021deeperthings, zhou2019adaptive}); (2) \textit{split-part}: a split-part is a partial layer-volume. Each (sub-)layer in a split-part is a partial layer in the layer-volume. The number of (sub-)layers in a split-part is equal to that in the layer-volume; (3) \textit{horizontal partition}: the division of a CNN model into layer-volumes; (4) \textit{vertical split}: the division of a layer-volume into split-parts; (5) \textit{partition location}: the original index (in the CNN model) of the first layer in a layer-volume; (6) \textit{partition scheme}: the set of partition locations in a CNN distribution strategy; (7) \textit{split decision}: a set of coordinate values on the height dimension of the last layer in the layer-volume where the layer-volume is vertically split (We focus on the split on one dimension in this paper and leave splitting on multiple dimensions in the future work); (8) \textit{service requester}: the device with CNN inference service requests; (9) \textit{service providers}: the devices that can provide computing resources for CNN inference.

\begin{figure}[!t]
  \centering
  \includegraphics[width=0.48\linewidth,height=1.55in]{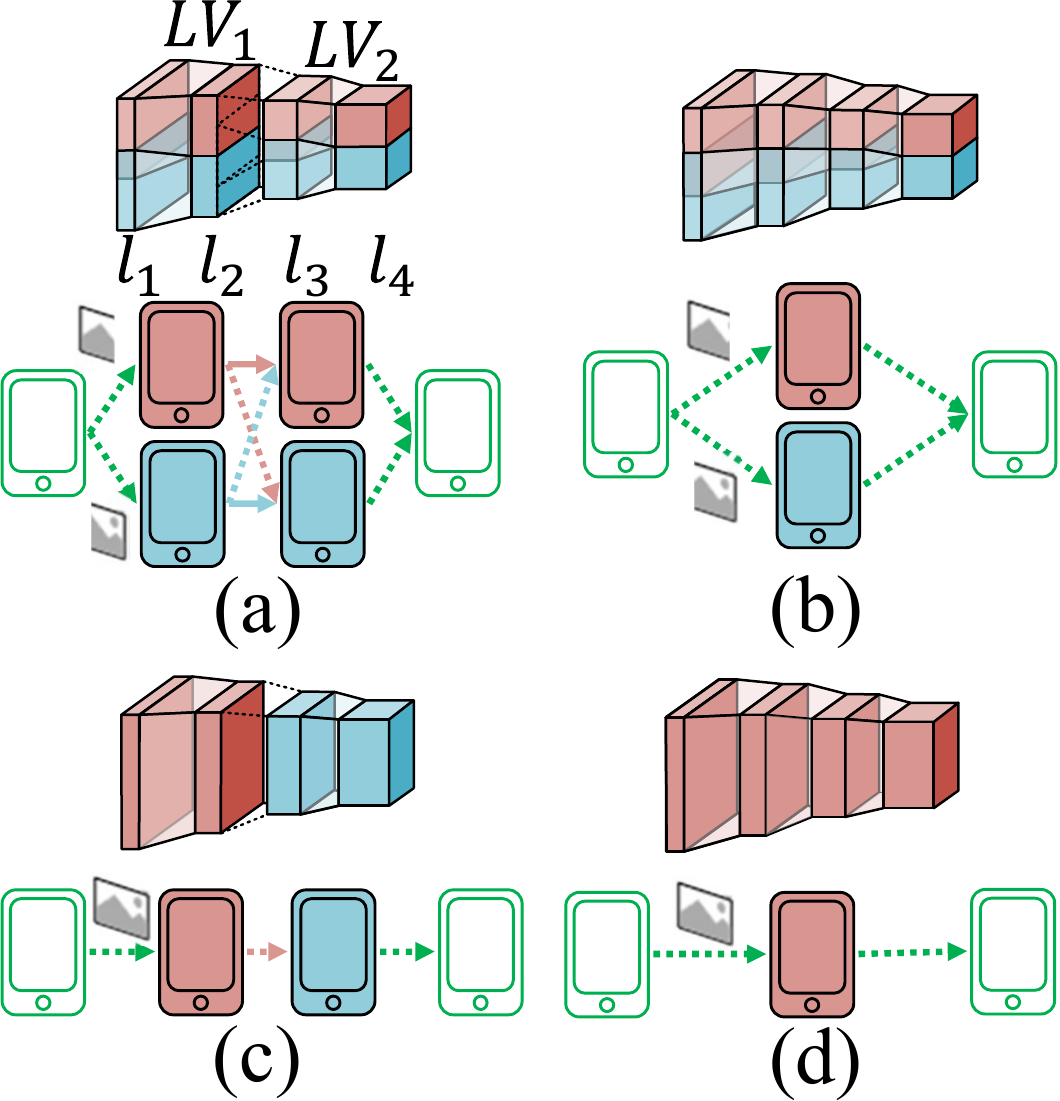}
  \vspace{-15pt}
  \vspace{0.1in}
  \caption{\small \textit{CNN inference distribution on two edge devices (blue and red one): (a) a general distribution form; (b) parallel distribution; (c) sequential distribution; (d) single-device offloading.}}\label{fig:cnn_device}
  \vspace{-0.2in}
\end{figure}

Fig.~\ref{fig:cnn_device} shows four examples of CNN inference distribution on two edge devices. Note that the distribution solutions in Fig.~\ref{fig:cnn_device} (b)(c)(d) are special cases of that in Fig.~\ref{fig:cnn_device} (a). Specifically, Fig.~\ref{fig:cnn_device} (b) takes the whole model as one layer-volume (i.e., parallel distribution on edge devices). Fig.~\ref{fig:cnn_device} (c) takes each layer-volume as one split-part (i.e., sequentially distribution on edge devices). Fig.~\ref{fig:cnn_device} (d) takes the whole model as one layer-volume and the layer-volume as one split-part (i.e., offloading to a single device). The design of DistrEdge naturally covers these special distribution forms.

\subsection{Vertical-Splitting Law}

A convolutional layer is defined by the following configurations: input width ($W_{in}$), input height ($H_{in}$), input depth ($C_{in}$), output depth ($C_{out}$), filter size ($F$), stride ($S$), padding ($P$), and an activation function. A maxpooling layer is defined by the following configurations: input width ($W_{in}$), input height ($H_{in}$), input depth ($C_{in}$), filter size ($F$), and stride ($S$). The input/output data size and the number of operations of a convolutional/maxpooling layer can be determined given these configurations~\cite{hadidi2019collaborative}. For two consecutively connected layers, the dimension of the former layer's output data matches the dimension of the later layer's input data.

For a layer-volume $\Tilde{\mathcal{V}}$ with $n$ layers $\{\Tilde{l}_1, \Tilde{l}_2, ..., \Tilde{l}_n\}$ ($n \geq 1$), a split-part $p_k$ is a partial layer-volume with $n$ sub-layers. Each sub-layer is a partial layer of each original layer, e.g., the $n$-th sub-layer of $p_k$ is a partial layer of $\Tilde{l}_n$. As we focus on splitting a layer-volume by the height dimension only, the output width and depth of $p_k$ are the same as those of $\Tilde{l}_n$. We denote the output height of $p_k$ as $h^{n,k}_{out}$. The output height of the $i$-th sub-layer is determined by the height of output data from the $(i+1)$-th sub-layer:
\begin{align}
    \vspace{-50pt}
    &h^{i,k}_{out} = (h^{i+1,k}_{out}-1)S_{i+1}+F_{i+1}, \label{eq:iheight}
\end{align}
where $S_{i+1}$ and $F_{i+1}$ are the stride and filter size of $\Tilde{l}_{i+1}$, respectively. Thus, given $h^{n,k}_{out}$, the output heights the first $(n-1)$ sub-layers can be determined by Eq.~\ref{eq:iheight} one by one. Further, the input height of $p_k$ can be calculated by:
\begin{align}
    &h^{1,k}_{in} = (h^{1,k}_{out}-1)S_1+F_1, \label{eq:1height}
\end{align}
The Vertical-Splitting Law is described as:

\noindent \textbf{Vertical-Splitting Law (VSL)}: For a split-part of a layer-volume with more than one layer, once the output height of its last sub-layer is determined, the input height of its first sub-layer can be calculated by Eq.~\ref{eq:iheight} and Eq.~\ref{eq:1height}.

\subsection{Challenges}

We summarize the following challenges in finding an optimal CNN inference distribution strategy on edge devices: \textit{1) High Dimensions of Search Space}: Most CNN models for computer vision tasks have more than ten layers (mainly convolutional layers), and the output size (height) from each layer ranges from less than $10$ to over $200$~\cite{simonyan2014very, he2016deep, szegedy2016rethinking, ren2015faster, redmon2016yolo9000}. It is infeasible to brute-force search over all possible combinations of distribution strategies. \textit{2) Heterogeneous and Nonlinear Device Characters}: In practice, the edge devices available for CNN inference distribution can be heterogeneous and with nonlinear characters. The split-parts need to be adjusted accordingly. However, as discussed in Section II, state-of-the-art distribution methods do not fully consider these characters when making their distribution strategies. \textit{3) Various Configurations across Layers in a CNN Model}: The layer configurations (e.g., height, depth, etc.) change across layers in a CNN model, which leads to various performance (i.e., latency). Consequently, the performance of distributed inference is also affected by the layer configurations. \textit{4) Dependency between Layers in CNN Model}: For most CNN models, the layers are connected sequentially one after another~\cite{simonyan2014very, he2016deep, szegedy2016rethinking, ren2015faster, redmon2016yolo9000}. While the operations of the same layer (layer-volume) can be computed in parallel, the operations of two layers (layer-volumes) cannot be computed in parallel. Thus, the devices' executions of the latter layer-volume are affected by the accumulated latencies of the former layer-volume. Consequently, when making splitting decisions to a layer-volume, it is important to consider the accumulated latencies of the split-parts in the former layer-volume.

\section{DistrEdge}
To utilize CNN inference distribution with DistrEdge, a computer (e.g., a laptop) collects information of the service requester and service providers and runs DistrEdge to find an optimal distribution strategy. The information collected includes network throughput between any two devices, layers' configurations of the CNN model, and CNN layer computation/transmission latency profiling on each provider. DistrEdge allows various forms to express the profiling results of a device. It can be regression models (e.g., linear regression, piece-wise linear regression, k-nearest-neighbor) or a measured data table of computing latencies with different layer configurations. Based on the optimal strategy, the controller informs the requester to send the split-parts (i.e., weights and biases) to the corresponding providers. The controller also informs the requester and providers to establish necessary network connections. Afterward, the CNN inference can be offloaded in a distributed manner on these service providers.

\subsection{Workflow of DistrEdge}
\begin{figure}[!ht]
  \centering
  \includegraphics[width=0.82\linewidth,height=2.5in]{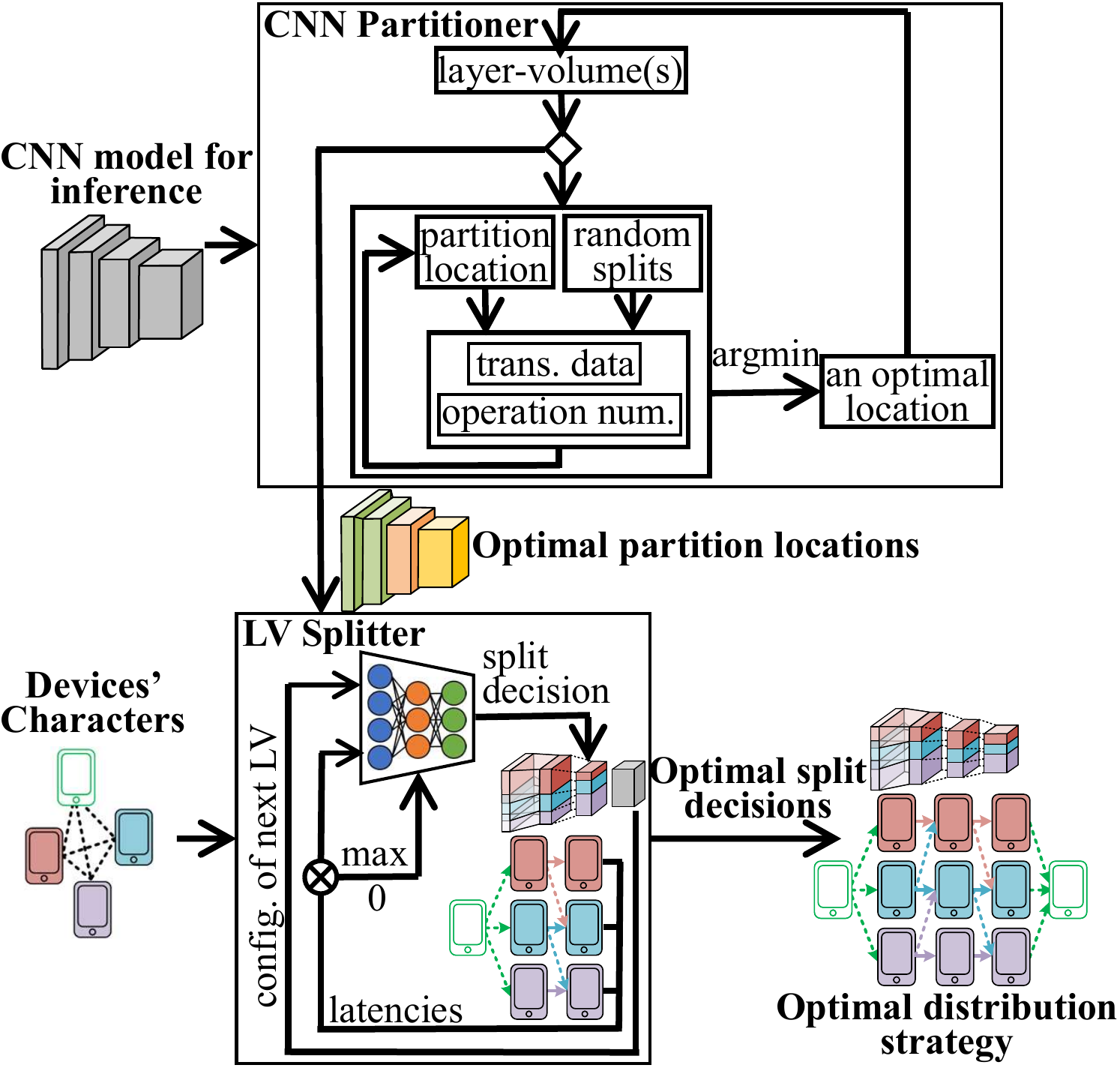}
  \vspace{-5pt}
  \caption{\small \textit{Workflow of DistrEdge.}}\label{fig:oss}
  \vspace{-0.2in}
\end{figure}

The workflow of DistrEdge is shown in Fig.~\ref{fig:oss}, which consists of two modules. The first module is a CNN partitioner that partitions a CNN model horizontally. The second module is a layer-volume splitter that splits layer-volumes vertically. The partitioner utilizes the layers' configurations of the CNN model to find the partition scheme based on the amount of transmission data and the number of operations. Given the partition scheme from the first module, the second module splits the layer-volumes with DRL. By modeling the split process as an MDP, a DRL agent is trained to make optimal split decisions for each layer-volume one by one based on the observations of accumulated latencies and layer configurations (the reward and states). The accumulated latencies implicitly reflect the device characters (i.e., computing capabilities and network conditions). During training, the latencies can be directly measured with real execution on devices or estimated by the profiling results. In this way, DistrEdge can form an inference distribution strategy on service providers.

\vspace{-7pt}

\subsection{Design of CNN Partitioner}
Algorithm~\ref{alg:hlc-pss} shows the proposed partitioner, Layer Configuration based Partition Scheme Search (LC-PSS). As the execution latency is affected by the amount of transmission data and the number of operations in CNN inference distribution~\cite{zhao2018deepthings, zhou2019adaptive}, we define a score $C_p$ as the function of operation number and transmission amount:
\begin{equation} \label{eq:partition}
    C_p = \alpha \cdot \mathcal{T} + (1 - \alpha) \cdot \mathcal{O}.
\end{equation}
where $\mathcal{O}$ and $\mathcal{T}$ are the total number of operations and the total amount of transmission data in the CNN model, respectively. The hyper-parameter $\alpha$ is used to control the trade-off between the two factors. Given a CNN model $\mathcal{M}$, a partition scheme $R_p$, and split decisions $R_s$, $\mathcal{O}$ and $\mathcal{T}$ can be determined. We generate a set of random split decisions $\mathbf{R}^r_s$. With $\mathbf{R}^r_s$, we average the scores of all the $C_p (R^i_s)$ ($R^i_s \in \mathbf{R}^r_s$) given $R_p$ and $\mathcal{M}$, denoted as $\Bar{C}_p$. We can formulate the optimal partition scheme search problem as:
\begin{equation} \label{eq:part_op}
    \min_{R_p} \frac{1}{|\mathbf{R}^r_s|}\sum_{i=1}^{|\mathbf{R}^r_s|} C_p (R_p, R^i_s, \mathcal{M}), R^i_s \in \mathbf{R}^r_s
\end{equation}
where $|\cdot|$ represents the number of elements (random split decisions) in $\mathbf{R}^r_s$, $C_p$ can be calculated by Eq.~\ref{eq:partition} given partition locations $R_p$ and split decisions $R^i_s$ of model $\mathcal{M}$. The LC-PSS greedily search for optimal $R_p$ that minimizes Eq.~\ref{eq:part_op}. In the $(k+1)$-th loop, the LC-PSS searches for an optimal partition location for each layer-volume obtained in the $k$-th loop (line 4 to 12). The optimal partition locations found in a loop is appended to $R^*_p$ (line 11). The loop stops when no more optimal partition locations are appended to $R^*_p$ (line 12, 13). In the worst case, there will be $(1+|M|)\cdot|M|/2$ loops to find the optimal partition strategy (i.e., when the optimal partition scheme is to partition the CNN model layer by layer). In contrast, if we brute-forcely search over all possible partition strategies for a CNN model $\mathcal{M}$~\cite{zhou2019adaptive}, there will be $(|M| \cdot |M-1| \cdot ... \cdot 2)$ loops. 

\begin{algorithm}
 \KwIn{$\mathcal{M}$: the CNN model for inference;
       \newline $\alpha$: the coefficient for partitioning score $C_p$;
       \newline $\mathbf{R}^r_s$: a set of random split decisions.
      }
 Initialize partitioning points record $R_p=\{1, |\mathcal{M}|\}$. \\
 \While{True}{
  $R_p^* \leftarrow R_p$; \\
  \For{$i=1, |R_p|-1$}{
   $C_p^*=\infty$; \\
   \For{$j=R_p[i], R_p[i+1]$}{
    $R'_p=R_p\cup\{j\}$; \\
    \small$\bar{C}_p=\frac{1}{|\mathbf{R}^r_s|}\sum_{i=1}^{|\mathbf{R}^r_s|} C_p (R'_p, R^i_s, \mathcal{M}), R^i_s \in \mathbf{R}^r_s$;\\
    \normalsize\If{$\bar{C}_p<C_p^*$}{
    $C_p^* = \bar{C}_p$;
    $j^* = j$;
    }
   }
  $R_p^* = R_p^* \cup \{j^*\}$; \\
  }
  \If{$|R_p|=|R_p^*|$}{
  break;}
  \Else{$R_p \leftarrow R_p^*$.}
 }
 \KwOut{$R_p^*$.}
\caption{\small{Layer Conf. based Partition Scheme Search}}\label{alg:hlc-pss}
\end{algorithm}

\subsection{Design of LV Splitter}

Given service providers $\mathcal{D}$, the CNN model $\mathcal{M}$, and the optimal partition scheme $R_p^*$, the LV splitter is to make optimal splitting decisions $R_s^*$ that minimizes the end-to-end-execution latency:
\begin{align}
    \min_{R_s} T(\{\mathcal{M}, R_p^*\}; \mathcal{D}; R_s) \label{eq:lvs}
\end{align}
where $R_s$ is a series of splitting decisions for all layer-volumes $\{\mathcal{V}_1, ..., \mathcal{V}_{L}\}$ of $\mathcal{M}$, $L$ is the number of layer-volumes in $\mathcal{M}$.

\subsubsection{Modeling as a Markov Decision Process}
The problem of Eq.~\ref{eq:lvs} can be modeled as an MDP. We define the \textit{action} $\boldsymbol{a}_l$ as the splitting parameters for a layer-volume $\mathcal{V}_l$:
\begin{align}
    \boldsymbol{a}_l = (x^{l}_1, x^{l}_2, ..., x^{l}_{|\mathcal{D}|-1}) \label{eq:action}
\end{align}
where $|\mathcal{D}|$ is the total number of service providers, $x^{l}_i$ $(i \in [1, |\mathcal{D}|-1])$ is the location on the height-dimension of the last layer in $\mathcal{V}_l$. The locations satisfy $x^{l}_i \in [0, H^{l}]$ and $x^{l}_i \leq x^{l}_j$ when $(i<j)$,
where $H^{l}$ is the height of the last layer in $\mathcal{V}_l$. The $i$-th split-part in $\mathcal{V}_l$ is allocated to the $i$-th service provider. The last sub-layer of the $i$-th split-part is a sub-layer with operations between $x^{l}_{i-1}$ and $x^{l}_{i}$ on the last layer of $\mathcal{V}_l$ ($x^{l}_{0}=0$ and $x^{l}_{|\mathcal{D}|}=H^{l}$). The dimension of the rest sub-layers of the $i$-th split-part can be determined by Eq.~\ref{eq:iheight}. The \textit{state} $\boldsymbol{s}_l$ is defined as:
\begin{align}
    \boldsymbol{s}_l = (\boldsymbol{T}_{l-1}, H^l, C^l, F^l, S^l)
    \label{eq:state}
\end{align}
where $\boldsymbol{T}_{l-1}$ are accumulated latencies on the service providers after computation of their allocated split-parts of $\mathcal{V}_{l-1}$, $(H^l, C^l, F^l, S^l)$ are configurations of the last layer in $\mathcal{V}_l$ (i.e., height, depth, filter size, and stride).

At step $l$, the action $\boldsymbol{a}_l$ determines the shapes of the split parts of $\mathcal{V}_l$ allocated to the service providers. On the one hand, the computing latency of a split-part is related to: (1) the shape of its split-part, (2) the device's characters, (3) the layer configurations. On the other hand, the accumulated latencies $\boldsymbol{T}_{l}$ are related to: (1) the computing latencies of the split parts in $\mathcal{V}_l$, (2) the transmission latencies occurred between (the split-parts of) $\mathcal{V}_{l-1}$ and $\mathcal{V}_l$, (3) the accumulated latencies $\boldsymbol{T}_{l-1}$ for $\mathcal{V}_{l-1}$. Therefore, the accumulated latencies $\boldsymbol{T}_{l}$ are determined by $\boldsymbol{a}_l$ and $\boldsymbol{s}_l$ together. As $\boldsymbol{T}_{l}$ are partial of $\boldsymbol{s}_{l+1}$, the splitting problem of Eq.~\ref{eq:lvs} can be regarded as an MDP. In other words, the next state $\boldsymbol{s}_{l+1}$ depends on the current state $\boldsymbol{s}_l$ and the action $\boldsymbol{a}_l$.

\subsubsection{Splitting Decision Search with DRL} Compared to other optimization algorithms, DRL can deal with the complex environment by learning and can find optimal actions efficiently without an analytical model of the environment. By modeling the split process as an MDP, we can utilize DRL to make optimal split decisions. As we aim at minimizing the end-to-end execution latency $T$ of $\mathcal{M}$, we define the reward $r_l$ as:
\begin{align} \label{eq:reward}
    \begin{cases}
        r_l = 0\ \ \ \ \ (l<L),\\
        r_l=1/T\ \ (l=L). 
    \end{cases}
\end{align}


For an layer-volume splitter, the splitting parameters (decisions) $(x^{l}_1, x^{l}_2, ..., x^{l}_{|\mathcal{D}|-1})$ are non-negative integers ranging from $0$ to $H^l$. However, due to this problem's high search space dimension, a discrete action space has an extremely high dimension. Moreover, as the configurations vary across layers in a CNN model, the dimension of a discrete action space changes across layer-volumes. Thus, it is impractical to utilize DRL algorithms with discrete action space in the LV splitter. When applying DRL algorithms with continuous action space, the \textit{action mapping function} is defined:
\begin{align} \label{eq:mf}
    x^l_i = round(H^l \cdot \frac{\Tilde{x}^l_i-A}{B-A}),\ (1 \leq i \leq |\mathcal{D}|-1)
\end{align}
where $round(\cdot)$ is to round a real value to an integer, $\Bar{x}^l_i$ is an element in $\mathbf{a}_l$, $[A, B]$ is the activation function' boundary in DRL agent. With Eq.~\ref{eq:state}, Eq.~\ref{eq:reward}, Eq.~\ref{eq:action}, and Eq.~\ref{eq:mf}, the Optimal Split Decision Search (OSDS) is described in Algorithm~\ref{alg:ddpg-osds}.

\begin{algorithm}
 \KwIn{$\{\mathcal{V}_1, ..., \mathcal{V}_{L}\}$: layer-volumes of $\mathcal{M}$;
 \newline $\mathcal{D}$: service providers;
 \newline $Max_{ep}$: the maximum training episode;
 \newline $\Delta \epsilon, \sigma^2$: hyper-parameters for exploration;
 \newline $N_b, \gamma$: hyper-parameters for training DRL.\\
 }
 Randomly initialize critic network $Critic(\boldsymbol{s}, \boldsymbol{a} | \theta_c)$ and actor network $Actor(\boldsymbol{s} | \theta_a)$ with parameters $\theta_c$ and $\theta_a$\;
 Initialize target networks $Critic'$ and $Actor'$ with parameters $\theta'_c \leftarrow \theta_c$ and $\theta'_a \leftarrow \theta_a$\;
 Initialize replay buffer $\mathcal{B}$.\\
 $T^* = \infty$\;
 \For{episode=$1$, $Max_{ep}$}{
  Initialize splitting decisions record $R_s=\{\}$\;
  Receive initial observation state $\boldsymbol{s}_1$\;
  $\epsilon = 1 - (episode \cdot \Delta \epsilon)^2$\;
  \For{h=$1$, $L$}{
  \If{$random<\epsilon$}{
  $\Tilde{\boldsymbol{a}}_l = Actor(\boldsymbol{s}_l | \theta_a) + \boldsymbol{\mathcal{N}}(0, \sigma^2)$\;}
  \Else{
  $\Tilde{\boldsymbol{a}}_l=Actor(\boldsymbol{s}_l | \theta_a)$\;
  }
  Map sorted $\Tilde{\boldsymbol{a}}_l$ to $\boldsymbol{a}_l$ (Eq.~\ref{eq:mf})\;
  
  Store $\boldsymbol{a}_l$ into $R_s$\;
  Split $\mathcal{V}_l$ with $\boldsymbol{a}_l$ and allocate split parts to $\mathcal{D}$\;
  Observe next state $\boldsymbol{s}_{l+1}$ and reward $r_l$\;
  Store transition $(\boldsymbol{s}_l, \Tilde{\boldsymbol{a}}_l, r_l, \boldsymbol{s}_{l+1})$ to $\mathcal{B}$\;
  Sample $N_b$ transitions $(\boldsymbol{s}_i, \boldsymbol{a}_i, r_i, \boldsymbol{s}_{i+1})$ from $\mathcal{B}$\;
  Set \small $y_i = r_i + \gamma Critic'(\boldsymbol{s}_{i+1}, Actor'(\boldsymbol{s}_{i+1}|\theta'_a)|\theta'_c)$\;
  \normalsize Critic loss: \small $\frac{1}{N_b} \sum_i (y_i-Critic(\boldsymbol{s}_i, \boldsymbol{a}_i | \theta_c))^2$\;
  \normalsize Update critic and actor networks.
  }
  Observe end-to-end execution latency $T$\;
  \If{$T<T^*$}{
  $R^*_s \leftarrow R_s$; $T^*=T$;\\
  $Critic^* \leftarrow Critic$; $Actor^* \leftarrow Actor$.\\
  }
 }
 \KwOut{$R^*_s, Critic^*, Actor^*$.}
\caption{Optimal Split Decision Search (OSDS)}\label{alg:ddpg-osds}
\end{algorithm}


In Algorithm~\ref{alg:ddpg-osds}, DDPG is an actor-critic DRL algorithm for continuous action space~\cite{lillicrap2015continuous}. The DRL agent is trained for $Max_{ep}$ episodes to make optimal policy (split decisions) (Line 5). At the $l$-th step, the DRL agent makes split decisions for $\mathcal{V}_l$ (Line 9), either by random exploration (Line 11) or by the actor network (Line 13). We sort the elements in the original output action vector $\Tilde{\boldsymbol{a}}_l$ (Line 11 or 13) from low to high and map the sorted vector to the true splitting parameters $\boldsymbol{a}_l$ with Eq.~\ref{eq:mf} (Line 14). The layer-volume $\mathcal{V}_l$ is split by $\boldsymbol{a}_l$, and the split-parts are allocated to the service providers in $\mathcal{D}$ (Line 16). Afterward, the next state $\boldsymbol{s}_{l+1}$ and the reward $r_l$ can be observed (Line 17) either by simulation or real execution. In each episode, the end-to-end execution latency $T$ can be measured (Line 23). The optimal splitting decisions $R^*_s$, $T^*$, $Critic^*$, and $Actor^*$ are stored (Line 24 to 26). The updating of the critic network and the actor network is illustrated in Line 20 to 22. We still utilize the original output action vector $\Tilde{\boldsymbol{a}}_l$ for training the networks (Line 18). $N_b$ tuples are randomly sampled from $\mathcal{B}$ (Line 18, 19) to train the actor-critic networks.

\section{Experiments}

In this section, we first explain how the testbed is built. Then we demonstrate seven baselines of state-of-the-art CNN inference distribution methods. For the evaluation, we first observe the effect of different $\alpha$ in LC-PSS in different device/network cases. Then we compare the performance (IPS: image-per-second) of DistrEdge with the seven baseline methods in various heterogeneous cases (Table~\ref{tab:heter_devices},~\ref{tab:heter_bw}) and large-scale cases (Table~\ref{tab:large_scale}). The hyper-parameters in DistrEdge are set as follows: (1) LC-PSS: $\alpha=0.75$, $|\mathbf{R}^r_s|=100$; (2) OSDS: $Max_{ep}=4000$, $\Delta \epsilon = 1/250$, $\sigma^2=0.1$ with four service providers ($\sigma^2=1$ with 16 service providers),  $N_b=64$, $\gamma=0.99$, learning rates of the $Actor$ and $Critic$ network as $10^{-4}$ and $10^{-3}$, respectively. The $Critic$ network is consisted of four fully-connected layers with dimensions of $\{400, 200, 100, 100\}$ and $Actor$ network is consisted of three fully-connected layers with dimensions of $\{400, 200, 100\}$.

\subsection{Test Setup}
\begin{figure}[!t]
  \centering
  \includegraphics[width=0.4\linewidth,height=1.2in]{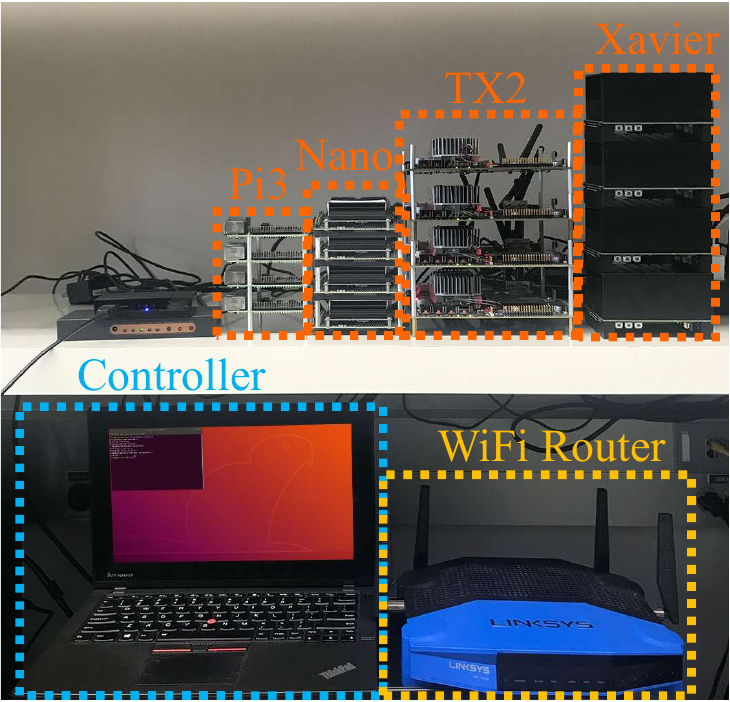}
  \vspace{-10pt}
  \caption{\small \textit{Testbed.}}\label{fig:testbed}
  \vspace{-0.3in}
\end{figure}

\begin{figure}[!ht]
  \centering
  \includegraphics[width=0.5\linewidth,height=1.3in]{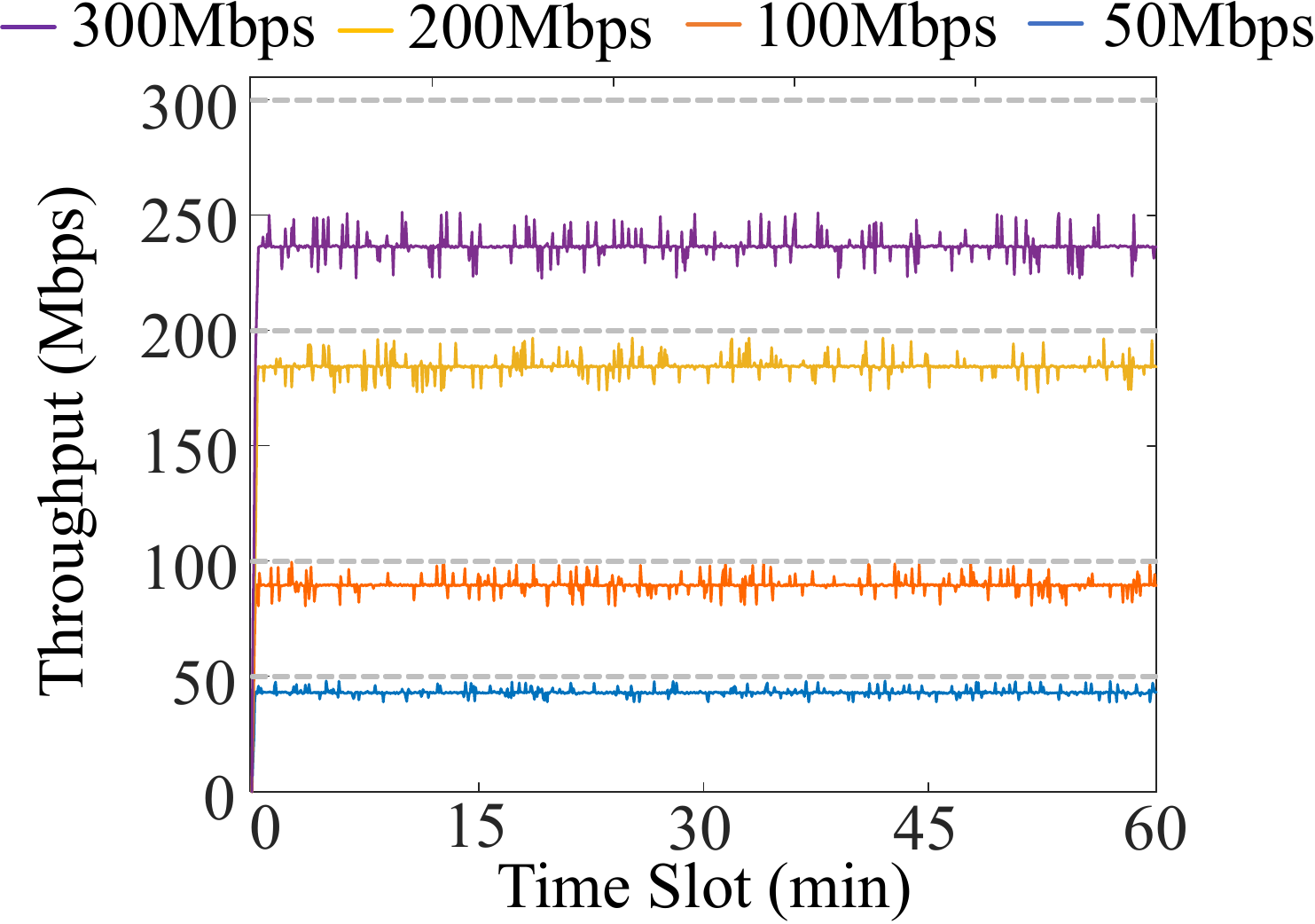}
  \caption{\small \textit{Sampled Network Throughput (Mbps) of WiFi w/ Different Bandwidth.}}\label{fig:wifi_bw}
  \vspace{-0.2in}
\end{figure}

Our testbed is shown in Fig.~\ref{fig:testbed}. Four types (Raspberry Pi3, Nvidia Nano, Nvidia TX2, Nvidia Xavier) of devices are used as service providers. A laptop works as the service controller. A mobile phone is used as the service requester. The wireless router (Linksys AC1900) provides 5GHz WiFi connections among the devices. The sampled throughput traces of wireless network with WiFi bandwidth ($300$Mbps, $200$Mbps, $100$Mbps, $50$Mbps) are shown in Fig.~\ref{fig:wifi_bw}. The router is equipped with OpenWrt~\cite{openwrt} system, which allows us to control bandwidth between different devices. The network connections between devices are built with TCP socket. The CNN layers are implemented in TensorRT SDK~\cite{TensorRT} using FP16 precision and batch size 1. One note is that TensorRT is not a necessary part of the CNN inference distribution implementation. We adopt it to show that DistrEdge is compatible with popular deep learning inference optimizers like TensorRT. For the DRL training in OSDS, we profile the computing latency on each type of device and the transmission latency in each bandwidth level against the height of each layer in a CNN model (granularity as $1$). Specifically, we use TensorRT Profiler~\cite{Profiler} to profile the computing latency. The transmission latency is measured from the time when the data are read from the computing unit (i.e., GPU or CPU) on the sending device to the time when the data are loaded to the memory on the receiving device (both transmission latency and I/O reading/writing latency are included). Each measurement point is repeated $100$ times, and we then compute the mean values as the profiled latencies. The profiling results are shared with the service controller, and the controller can utilize the results whenever it needs to run OSDS.

To evaluate the performance of a CNN inference distribution strategy, we stream $5000$ images from the service requester to the service providers. An image will not be sent until the result of its previous image is received by the service requester. We measure the overall latency in processing the $5000$ images and compute averaged FPS. To reduce overhead, the images are split beforehand according to the distribution strategy. The split-parts on the providers are also preloaded to their memory. On each service provider, three threads are running parallel to implement computation, data receiving, and data transmission by sharing data with a queue. All the codes are implemented in Python3. We mainly focus on the optimal distribution of the convolutional layers and maxpooling layers in this paper. The last fully-connected layer(s) are computed on the service provider with the highest allocated amount of the last layer-volume in our experiments.

\subsection{Baseline Methods}

To evaluate the performance of DistrEdge compared to state-of-the-art CNN inference distribution methods, we take seven methods as our baseline: \textit{(1) CoEdge}~\cite{zeng2020coedge} (linear models for devices and networks, layer-by-layer split); \textit{(2) MoDNN}~\cite{mao2017modnn} (linear models for devices, layer-by-layer split); \textit{(3) MeDNN}~\cite{mao2017mednn} (linear models for devices, layer-by-layer split); \textit{(4) DeepThings}~\cite{zhao2018deepthings} (equal-split, one fused-layers); \textit{(5) DeeperThings}~\cite{stahl2021deeperthings} (equal-split, multiple fused-layers); \textit{(6) AOFL}~\cite{zhou2019adaptive} (linear models for devices and networks, multiple fused layers); \textit{(7) Offload}: We select the service provider with the best computing hardware (e.g., the best GPU) to offload the CNN inference. Note that equal-split refers to splitting layer(s) equally (DeepThings and DeeperThings). In contrast, CoEdge, AOFL, MoDNN, and MeDNN split layer(s) based on linear models of computation and/or transmission. Layer-by-layer refers to making split decisions per layer (CoEdge, MoDNN, and MeDNN). One fused-layer refers to fusing multiple layers once in a CNN model (DeepThings). Multiple fused-layers refer to allowing more than one fused-layers in a CNN model (DeeperThings and AOFL). The term \textit{fused-layer} is equivalent to \textit{layer-volume} in DistrEdge.

\subsection{$\alpha$ and $|\mathbf{R}^r_s|$ in LC-PSS}
\begin{figure}[!ht]
  \centering
  \includegraphics[width=1\linewidth,height=1in]{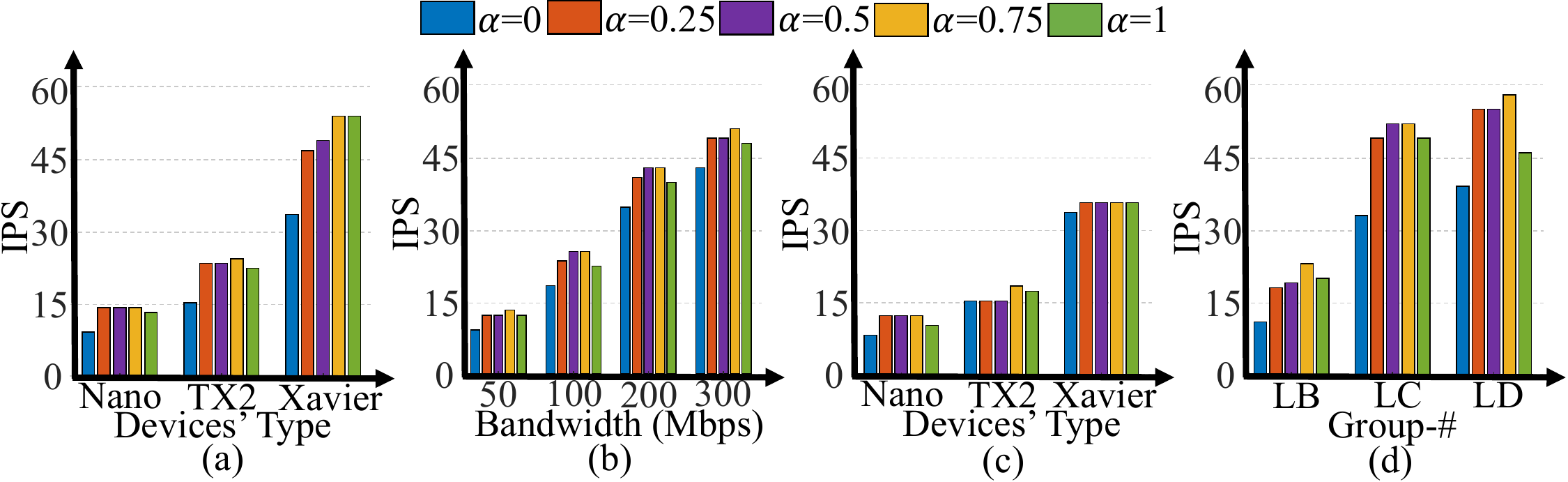}
  \vspace{-0.25in}
  \caption{\small \textit{IPS w/ different $\alpha$ (VGG-16): (a) homogeneous devices ($200$Mbp); (b) heterogeneous devices' types (Group DB); (c) heterogeneous network bandwidth (Group NA); (d) large-scale devices.}}\label{fig:alpha}
\end{figure}

\begin{figure}[!ht]
  \centering
  \vspace{-0.1in}
  \includegraphics[width=0.9\linewidth,height=1.25in]{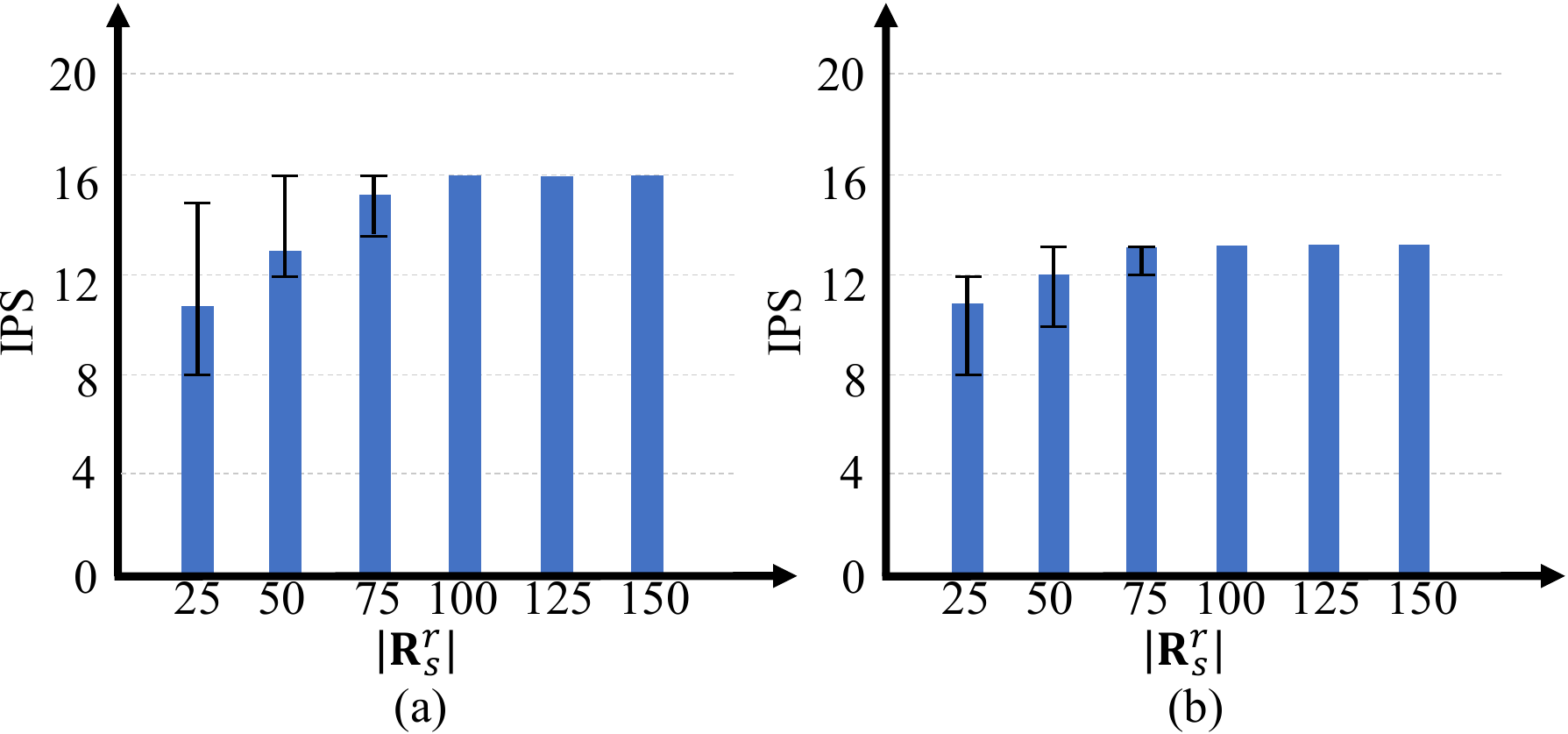}
  \vspace{-0.15in}
  \caption{\small \textit{Image per Second w/ different number of random split decisions in LC-PSS (VGG-16): (a) DB, 50Mbps WiFi; (b) NA, Nano.}}\label{fig:num_dec}
  \vspace{-0.2in}
\end{figure}

We first study the effect of the hyper-parameter $\alpha$ and $|\mathbf{R}^r_s|$ in LC-PSS (Algorithm~\ref{alg:hlc-pss}), as shown in Fig.~\ref{fig:alpha} and~\ref{fig:num_dec}. 

$\alpha$ controls the trade-off between $\mathcal{O}$ and $\mathcal{T}$ when calculating $C_p$ (Eq.~\ref{eq:partition}). We observe that, when $\alpha$ is smaller, LC-PSS tends to partition the CNN model into more layer-volumes, and vice versa. For example, VGG-16 is partitioned into $16$ layer-volumes with $\alpha=0$ and is partitioned into only two layer-volumes with $\alpha=1$. To observe the overall performance with different $\alpha$, we further run the OSDS (Algorithm~\ref{alg:ddpg-osds}) with the partition schemes obtained with $\alpha=\{0, 0.25, 0.5, 0.75, 1.0\}$ under four types of environments for VGG-16 inference: (1) We have four homogeneous service providers under $200$Mbps WiFi; (2) The service providers' types are heterogeneous (Group DB in Table~\ref{tab:heter_devices}); (3) The service providers' WiFi bandwidths are heterogeneous (Group NA in Table~\ref{tab:heter_bw}); (4) We have $16$ service (Group LB, LC, LD in Table~\ref{tab:large_scale}). The performance comparison is shown in Fig.~\ref{fig:alpha} (a) to (d). We can observe poor performance with $\alpha=0$ ($C_p$ is only affected by the number of operations) and $\alpha=1$ ($C_p$ is only affected by the amount of transmission data). The highest performance is achieved with $\alpha=0.75$ in all cases. Overall, we should consider both the number of operations and the amount of transmission data when searching for the optimal partition scheme. We set $\alpha=0.75$ in the rest of our experiments. This study also shows that distribution CNN model layer-by-layer~\cite{zeng2020coedge, mao2017modnn, mao2017mednn} is not efficient (e.g., $\alpha=0$).

$|\mathbf{R}^r_s|$ is the number of random split decisions in Eq.~\ref{eq:part_op}. It affects the averaged score $\bar{C}_p$ in Algorithm~\ref{alg:hlc-pss} (line 8). With a small $|\mathbf{R}^r_s|$, the optimal partition locations generated by LC-PSS can change significantly, as different groups of random split decisions lead to different $\bar{C}_p$. With a large $|\mathbf{R}^r_s|$, the optimal partition locations generated by LC-PSS does not show much difference, as different groups of random split decisions lead to similar $\bar{C}_p$. We compare the overall performance (IPS) with different $|\mathbf{R}^r_s|$ in two cases (Group-DB with 50Mbps WiFi and Group-NA with Nano), as shown in Fig~\ref{fig:num_dec}. For each $|\mathbf{R}^r_s|$ value, we repeatedly run LC-PSS $50$ times and record the optimal partition locations of each time. We aggregate the IPS performance over $50$ times for each $|\mathbf{R}^r_s|$ value in Fig.~\ref{fig:num_dec}, and the maximum, average, and minimum IPS among $50$ results are shown. When $|\mathbf{R}^r_s|$ is small (e.g., $25$ and $50$), the IPS varies in a wide range. When $|\mathbf{R}^r_s|$ is large (larger than or equal to $100$), the IPS keeps stable. Thus, we set $|\mathbf{R}^r_s|$ to $100$ in the rest of our tests.

\subsection{Performance under Heterogeneous Environments}

To evaluate the performance of DistrEdge in a heterogeneous environment, we built up multiple cases in the laboratory. These cases can be divided into three types: (1) cases with heterogeneous devices (Table~\ref{tab:heter_devices}); (2) cases with heterogeneous network bandwidths (Table~\ref{tab:heter_bw}); (3) cases with large-scale devices (Table~\ref{tab:large_scale}). We utilize image-per-second (IPS) as the performance metric.

\begin{table}[!htp]
    \centering
    \setlength\abovecaptionskip{0pt}
    \renewcommand{\arraystretch}{0.7}
    \vspace{-0.15in}
    \caption{Groups of Heterogeneous Devices' Types}
    \begin{tabular}{c|c}
    \hline
    Group \# & Devices' Types \\
    \hline
    DA & TX2$\times 2 +$Nano$\times 2$ \\
    \hline
    DB & Xavier$\times 2 +$Nano$\times 2$ \\
    \hline
    DC & Xavier$\times 1 +$TX2$\times 1 +$Nano$\times 1 +$Pi3$\times 1$ \\
    \hline
    \end{tabular}
    \label{tab:heter_devices}
    \vspace{-0.2in}
\end{table}

\begin{table}[!htp]
    \centering
    \setlength\abovecaptionskip{0pt}
    \renewcommand{\arraystretch}{0.7}
    \caption{Groups of Heterogeneous Network Bandwidths}
    \begin{tabular}{c|c}
    \hline
    Group \# & Network Bandwidths (Mbps) \\
    \hline
    NA & 50$\times 2 +$200$\times 2$ \\
    \hline
    NB & 100$\times 2 +$200$\times 2$ \\
    \hline
    NC & 200$\times 2 +$300$\times 2$ \\
    \hline
    ND & 50$\times 1 +$100$\times 1 +$200$\times 1 +$ 300$\times 1$ \\
    \hline
    \end{tabular}
    \label{tab:heter_bw}
    \vspace{-0.1in}
\end{table}

\begin{table}[t]
    \centering
    \setlength\abovecaptionskip{0pt}
    \renewcommand{\arraystretch}{0.7}
    \caption{Groups of Large-Scale Service Providers (16 Devices)}
    \begin{tabular}{c|c}
    \hline
    Case \# & \{(Bandwidths (Mbps), Devices' Types)\} \\
    \hline
    LA & \{(300, Nano), (200, Nano), (100, Nano), (50, Nano)\} $\times$ 4 \\
    \hline
    LB & \{(300, Pi3), (200, Nano), (100, TX2), (50, Xavier)\} $\times$ 4 \\
    \hline
    LC & \{(200, Pi3), (200, Nano), (200, TX2), (200, Xavier)\} $\times$ 4 \\
    \hline
    LD & \{(50, Pi3), (100, Nano), (200, TX2), (300, Xavier)\} $\times$ 4 \\
    \hline
    \end{tabular}
    \label{tab:large_scale}
    \vspace{-0.2in}
\end{table}

\subsubsection{Heterogeneous Devices} 
\begin{figure}[!ht]
  \centering
  \includegraphics[width=0.95\linewidth,height=1.25in]{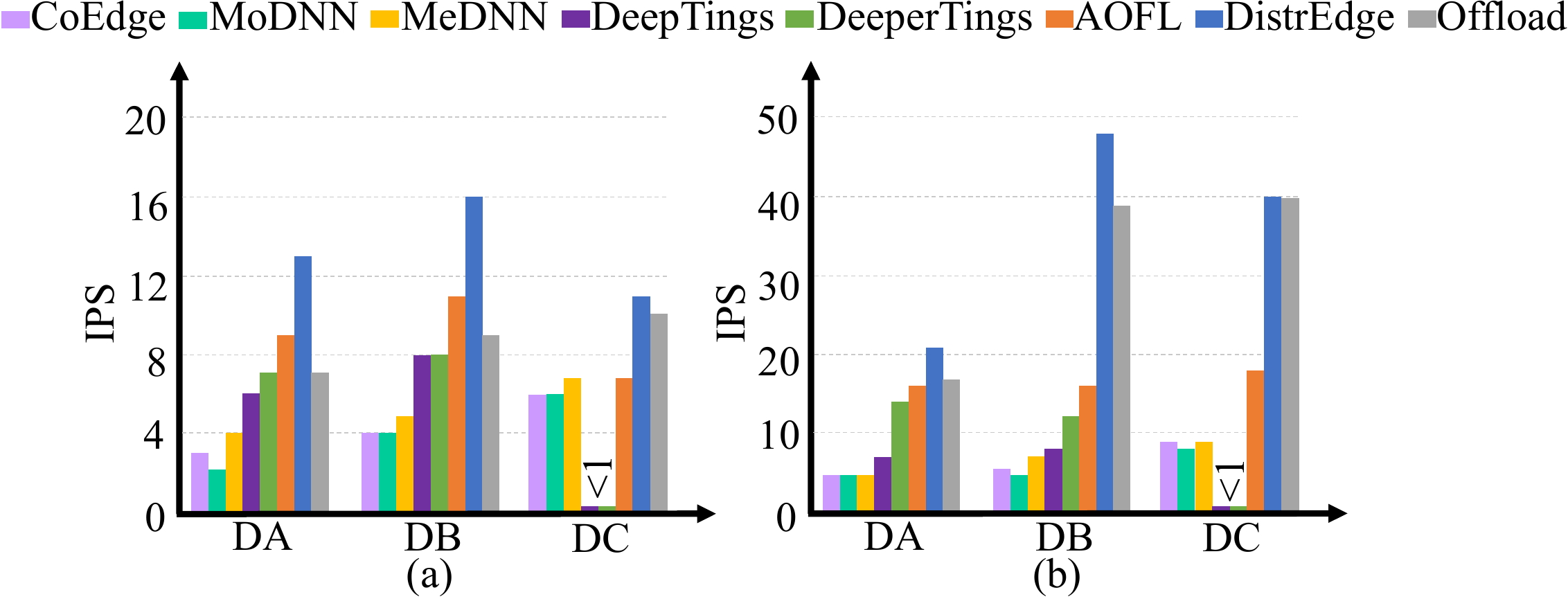}
  \vspace{-0.1in}
  \caption{\small \textit{Image per Second under Environments w/ Heterogeneous Devices (VGG-16): (a) 50Mbps WiFi; (b) 300Mbps WiFi.}}\label{fig:device_heter}
  \vspace{-0.1in}
\end{figure}

The performance of DistrEdge and baseline methods is shown in Fig.~\ref{fig:device_heter}. For each group, we set two network bandwidths \{50Mbps, 300Mbps\}. As shown in Fig.~\ref{fig:device_heter}, DistrEdge shows higher performance compared to the baseline methods in all cases of heterogeneous devices' types. The performance of DistrEdge shows higher performance improvement over the baseline methods in Group-DB, and DC (e.g., 1.5 to 3$\times$ over AOFL) compared to that in Group-DA (e.g., 1.2 to 1.5$\times$ over AOFL). As the computing capability comparison of the four devices' types is Pi3$<<$Nano$<$TX2$<$Xavier~\cite{nanovspi3, jetsonbenchmark}, the cases in Group-DB and Group-DC can be regarded as environments with higher heterogeneous levels compared to those in Group-DA. We can see that DistrEdge is more adaptive to heterogeneous devices' types compared to state-of-the-art methods. 

\subsubsection{Heterogeneous Network Bandwidths} 
\begin{figure}[!ht]
  \centering
  \vspace{-0.11in}
  \includegraphics[width=0.95\linewidth,height=1.25in]{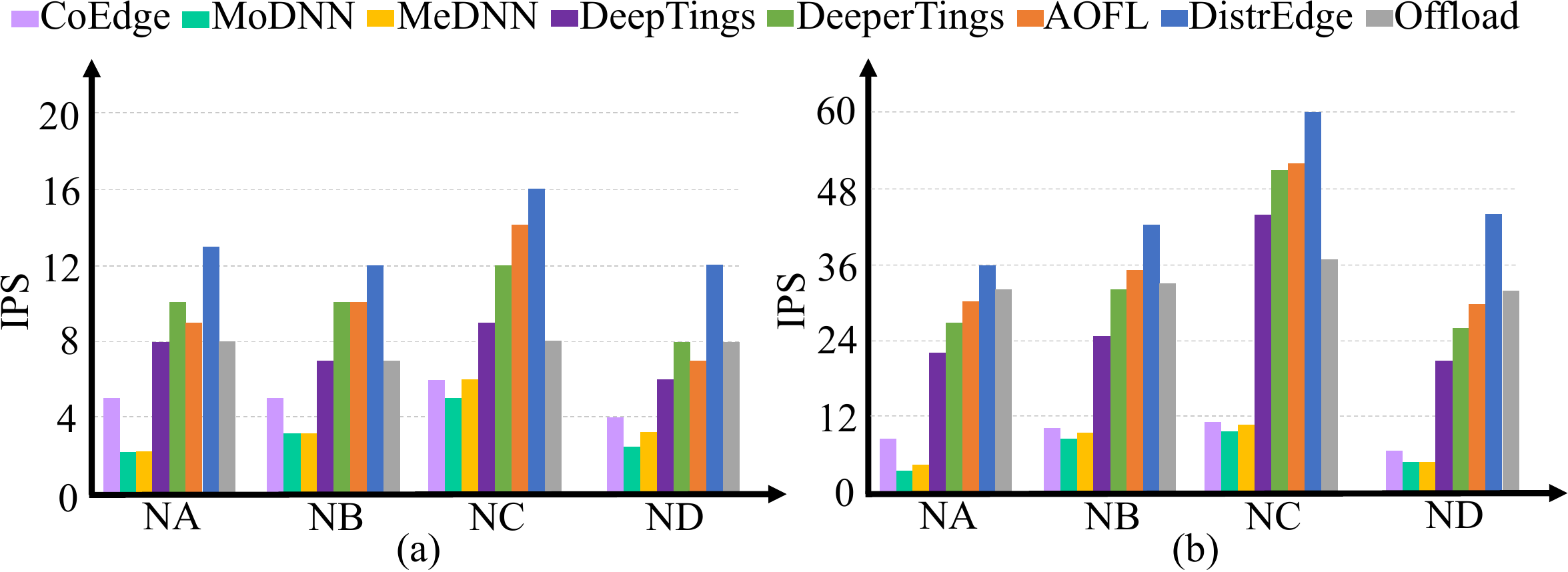}
  \vspace{-0.15in}
  \caption{\small \textit{Image per Second under Environments w/ Heterogeneous Networks (VGG-16): (a) Nano; (b) Xavier.}}\label{fig:network_heter}
  \vspace{-0.15in}
\end{figure}

The performance of DistrEdge and baseline methods is shown in Fig.~\ref{fig:network_heter}. For each group, we set two devices' types \{Nano, Xavier\}. As shown in Fig.~\ref{fig:network_heter}, DistrEdge shows higher performance compared to the baseline methods in all cases of heterogeneous network bandwidths. The performance of DistrEdge shows higher performance improvement over the baseline methods in Group-NA, and ND (e.g., 1.2 to 1.7$\times$ over AOFL) compared to that in Group-NB and NC (e.g., 1.1 to 1.3$\times$ over AOFL). As shown in Table~\ref{tab:heter_bw}, the cases in Group-NA and Group-ND are environments with higher heterogeneous levels compared to those in Group-NB and Group-NC. We can see that DistrEdge is more adaptive to heterogeneous network conditions compared to state-of-the-art methods.

\subsubsection{Large-Scale Devices} 
\begin{figure}[!ht]
  \centering
  \includegraphics[width=0.5\linewidth,height=1.2in]{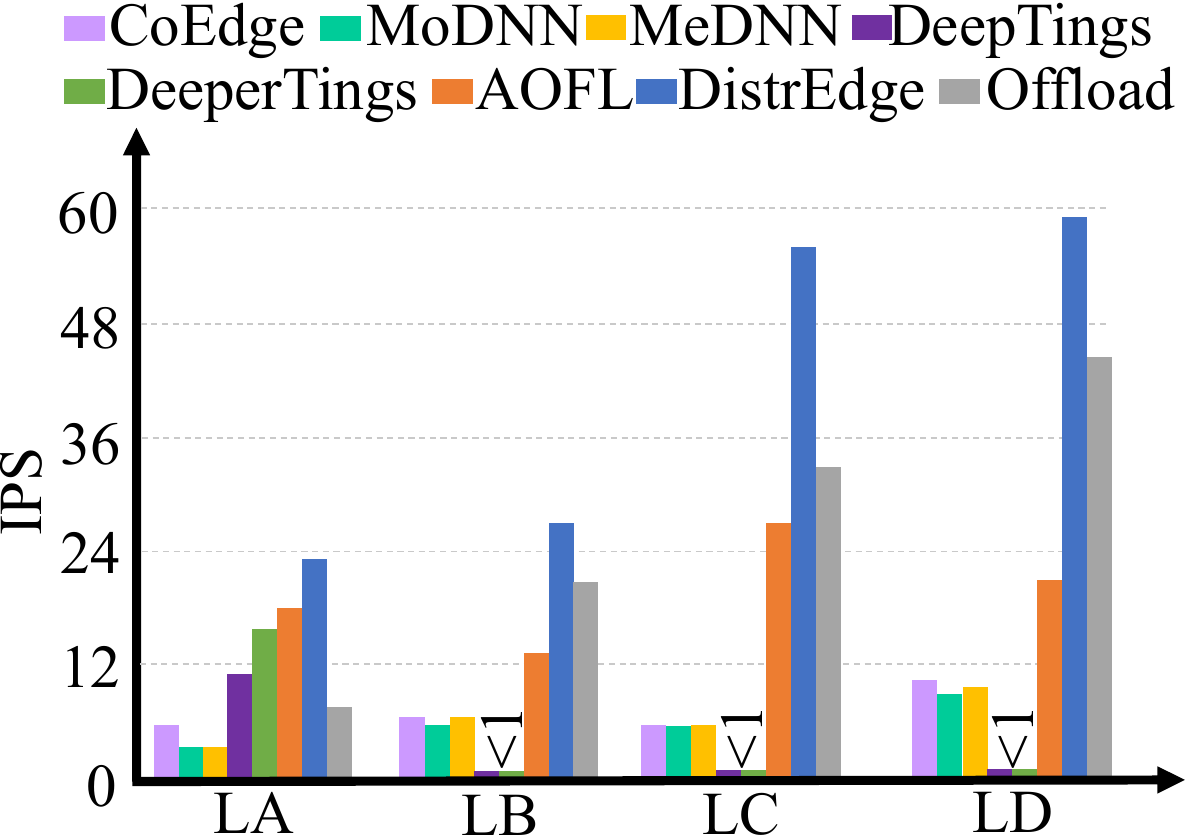}
  \vspace{-0.1in}
  \caption{\small \textit{Image per Second w/ Large-Scale Devices (VGG-16).}}\label{fig:large}
  \vspace{-0.15in}
\end{figure}

The performance of DistrEdge and baseline methods is shown in Fig.~\ref{fig:large}. Case-LA is with heterogeneous network bandwidth. Case-LB and Case-LD are with heterogeneous network bandwidth and devices' types. Case-LC is with heterogeneous devices' types. As shown in Fig.~\ref{fig:large}, DistrEdge shows higher performance in the four cases of large-scale service providers compared to the baselines (e.g., 1.3 to 3$\times$ over AOFL).

\subsection{Performance w/ Different Models}
\begin{figure}[!ht]
  \centering
  \includegraphics[width=0.9\linewidth,height=1.2in]{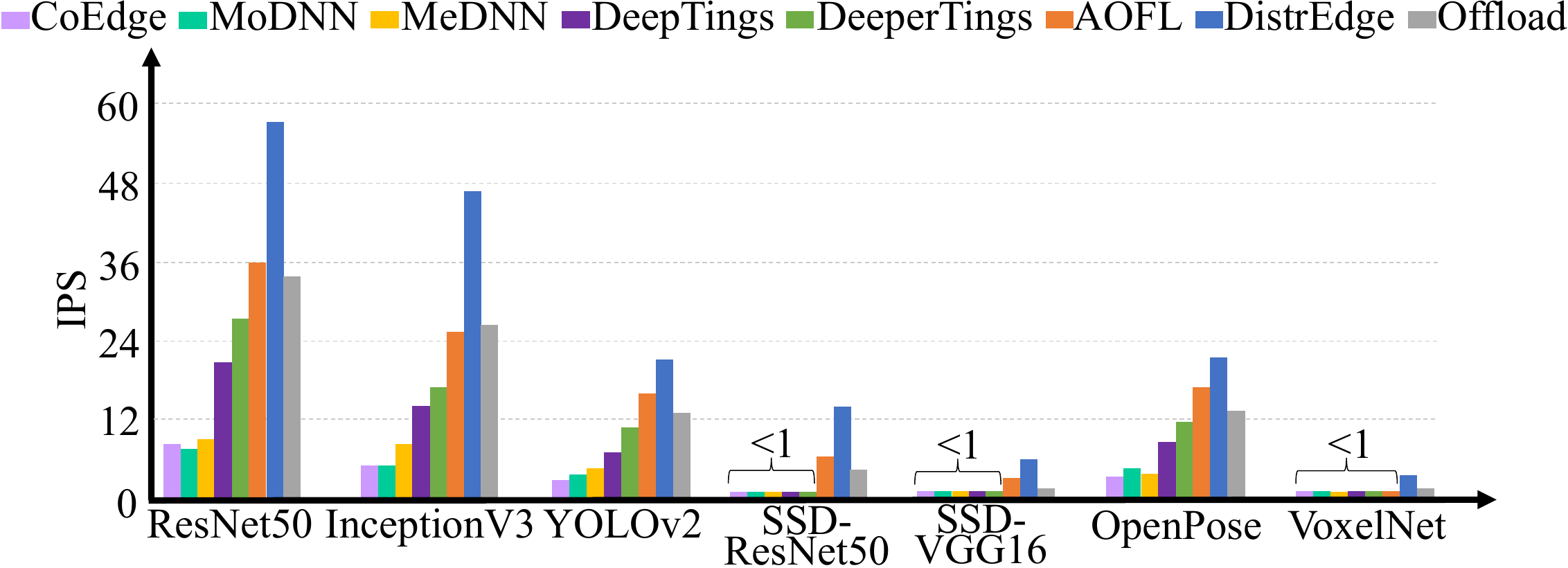}
  \vspace{-0.15in}
  \caption{\small \textit{Image per Second w/ Different Models (DB, 50Mbps).}}\label{fig:db_50}
  \vspace{-0.1in}
\end{figure}

\begin{figure}[!ht]
  \centering
  \includegraphics[width=0.9\linewidth,height=1.2in]{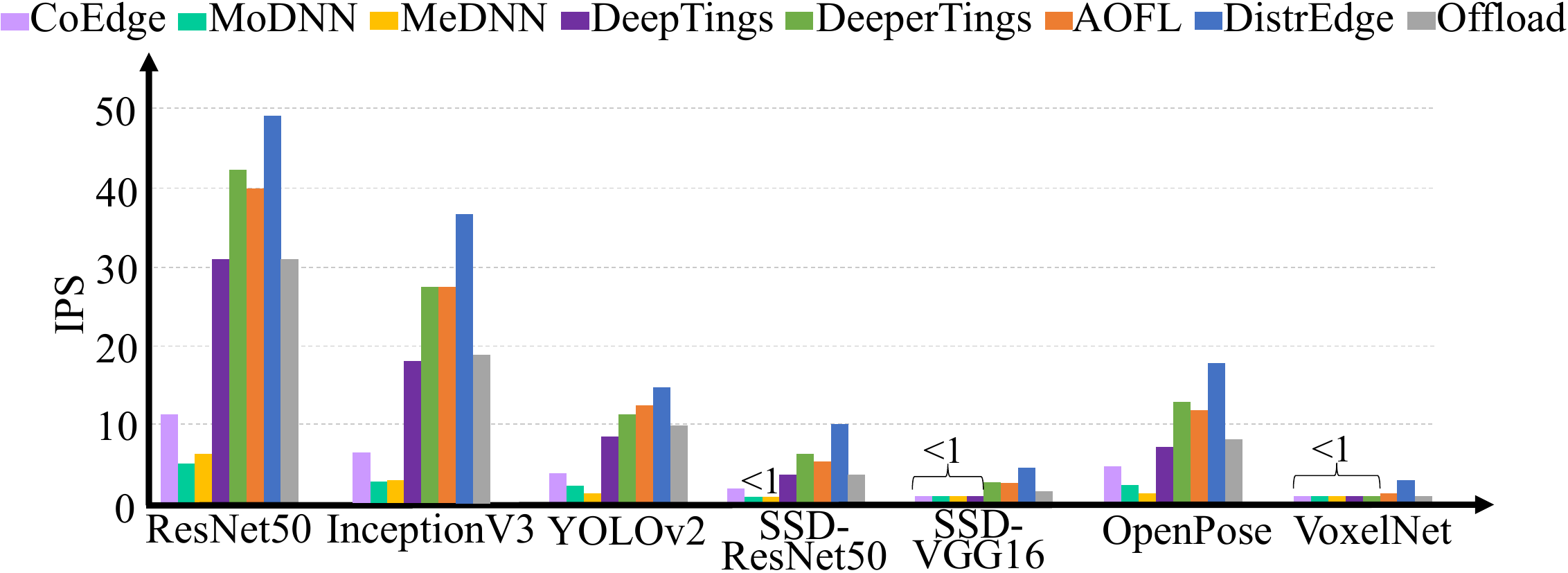}
  \vspace{-0.15in}
  \caption{\small \textit{Image per Second w/ Different Models (NA, Nano).}}\label{fig:na_nano}
  \vspace{-0.2in}
\end{figure}

Besides VGG-16 (classification), we implement inference distribution for other seven models of different applications: ResNet50~\cite{he2016deep} (classification), InceptionV3~\cite{szegedy2016rethinking} (classification), YOLOv2~\cite{redmon2016yolo9000} (object detection), SSD-ResNet50~\cite{liu2016ssd} (object detection), SSD-VGG16~\cite{liu2016ssd} (object detection), OpenPose~\cite{cao2019openpose} (pose detection), VoxelNet~\cite{zhou2018voxelnet} (3D object detection). We test under two cases: (1) Group-DB (Table~\ref{tab:heter_devices}) with 50Mbps WiFi, as shown in Fig.~\ref{fig:db_50}; (2) Group-NA (Table~\ref{tab:heter_bw}) with Nano, as shown in Fig.~\ref{fig:na_nano}. As shown in Fig.~\ref{fig:db_50} and Fig.~\ref{fig:na_nano}, DistrEdge outperforms the baseline methods in all the models by $1.1 \times$ to over $2.6 \times$.

\subsection{Performance in Highly Dynamic Networks}\label{sec:dynamic}
\begin{figure}[!ht]
  \centering
  \includegraphics[width=0.95\linewidth,height=1.25in]{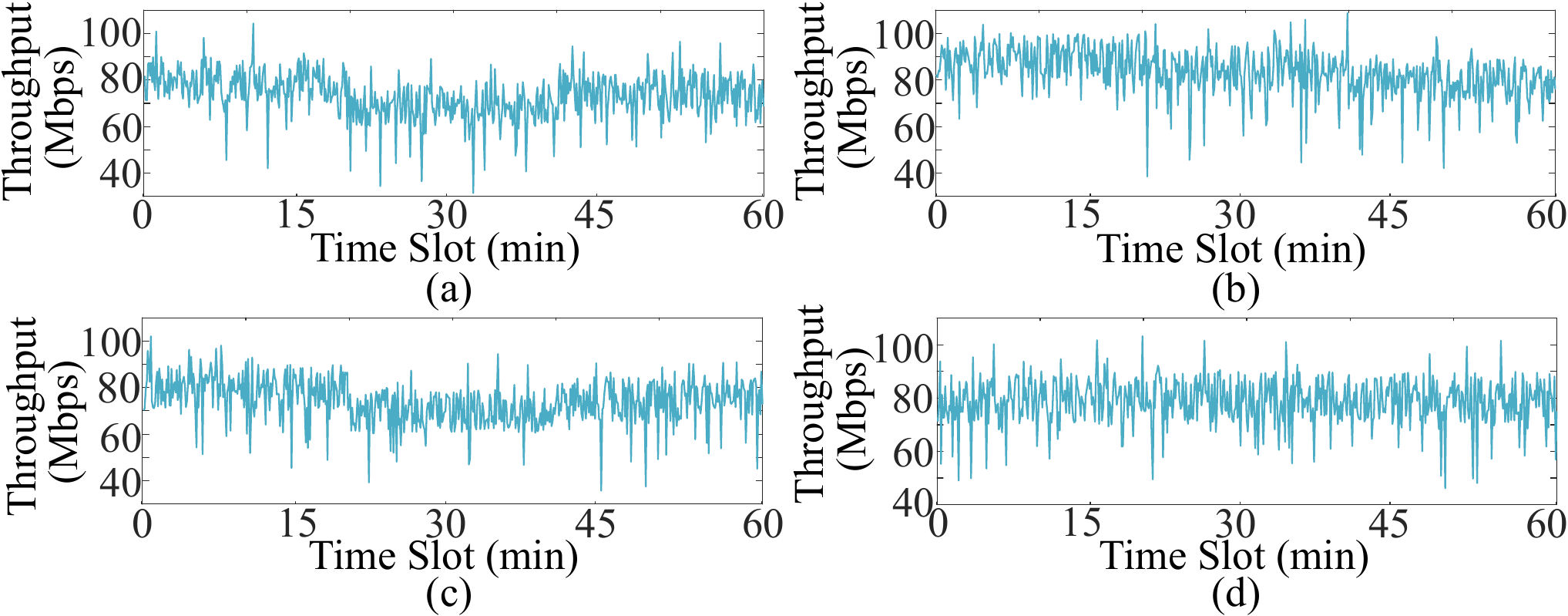}
  \vspace{-0.15in}
  \caption{\small \textit{Sampled Network Throughput (Mbps) of Highly Dynamic Network: (a) Device 1; (b) Device 2; (c) Device 3; (d) Device 4.}}\label{fig:lte_bw}
  \vspace{-0.15in}
\end{figure}

\begin{figure}[!ht]
  \centering
  \includegraphics[width=0.40\linewidth,height=1.1in]{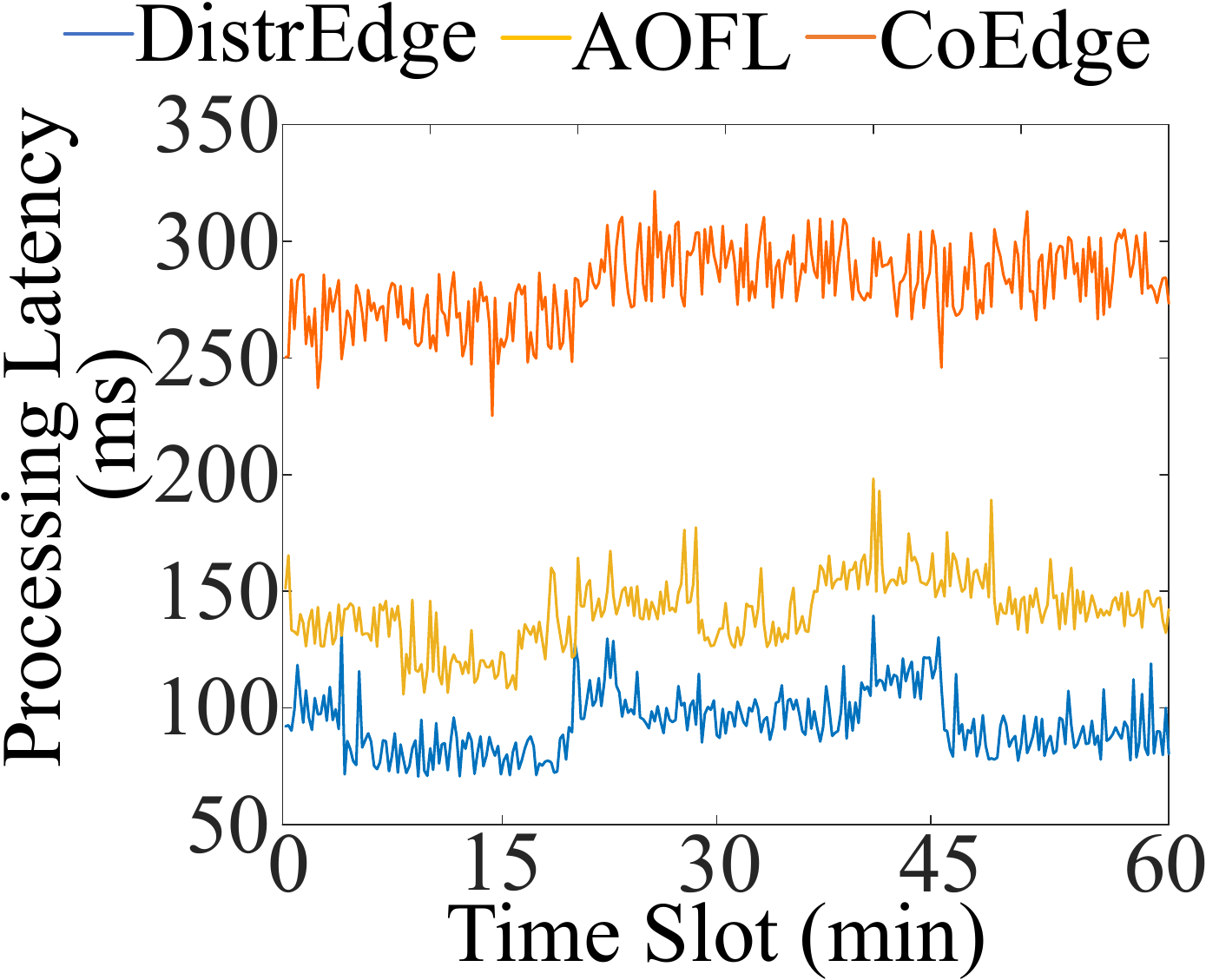}
  \vspace{-0.15in}
  \caption{\small \textit{Per-Image Processing Latency (ms) w/ Different Methods.}}\label{fig:lte_latency}
  \vspace{-0.2in}
\end{figure}

The above results are in the WiFi network which has small fluctuation shown in Fig.~\ref{fig:wifi_bw}. We further evaluate the performance of DistrEdge under highly dynamic network conditions. Specifically, we test with distribution on four devices (Nano) under highly dynamic network conditions. We generate four network throughput traces with high fluctuations shown in Fig.~\ref{fig:lte_bw}. Among the baseline methods, CoEdge~\cite{zeng2020coedge} and AOFL~\cite{zhou2019adaptive} are applicable to dynamic networks. We compare the performance of DistrEdge with CoEdge and AOFL, as shown in Fig.~\ref{fig:lte_latency}. The split decisions of all the three methods are made on a controller (ThinkPad L13 Gen 2 Intel) in an online manner based on the monitored network throughput (CoEdge and AOFL) and intermediate latency (DistrEdge). Note that the trained actor network of DistrEdge runs on the controller to make online split decisions. For DistrEdge and AOFL, the model partition locations are updated when averaged network throughput is detected to be changed significantly (e.g., at 20min and 40min time-slot in Fig.~\ref{fig:lte_bw}). For DistrEdge, the actor network is finetuned on the controller after partition location adjustment, which takes 20s to 210s in total. For AOFL, due to brute-force search for optimal partition locations, it takes 10min to update distribution scheme on the same controller.

As shown in Fig.~\ref{fig:lte_latency}, the per-image processing latency of CoEdge~\cite{zeng2020coedge} is the largest because it does not utilize layer-volumes to fuse layers. As data transmission occurs between any two connected layers, large transmission latency is generated, increasing the end-to-end execution latency of the model. In contrast, both AOFL~\cite{zhou2019adaptive} and DistrEdge reduce the transmission latency by fusing layers into layer-volumes. The key difference between AOFL and DistrEdge is that: AOFL makes split decisions based on a linear ratio, but DistrEdge makes split decisions based on the intermediate latency and learned relationship between latency and split decisions. As shown in Fig.~\ref{fig:lte_latency}, the per-image processing latency of DistrEdge is $40\%$ to $65\%$ of that of AOFL.

\subsection{Why DistrEdge Outperforms Baselines}

\begin{figure}[!ht]
  \centering
  \vspace{-0.1in}
  \includegraphics[width=0.6\linewidth,height=0.7in]{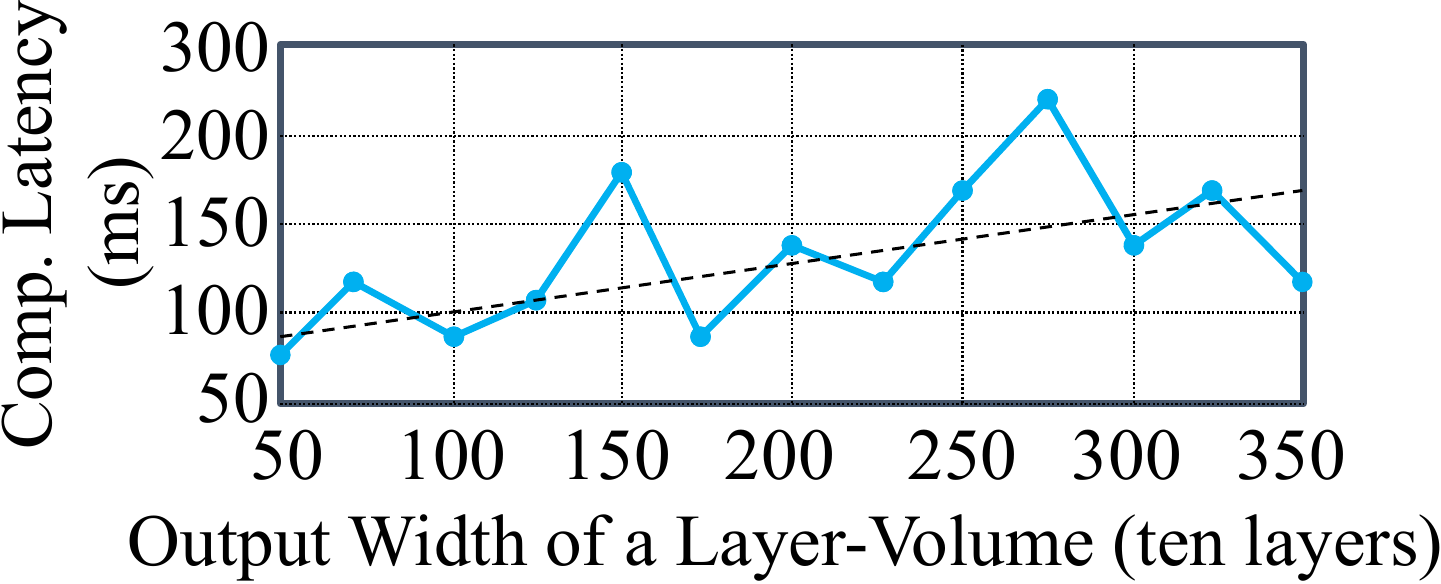}
  \vspace{-0.1in}
  \caption{\small \textit{Computing Latency v.s. Output Width in CNN.}}\label{fig:nonlinear}
  \vspace{-0.1in}
\end{figure}

For CoEdge~\cite{zeng2020coedge}, MoDNN~\cite{mao2017modnn}, and MeDNN~\cite{mao2017mednn}, as they split a CNN model layer-by-layer, data transmission through networks occurs between any two connected layers. Thus, they show large end-to-end latency due to transmission delay. For DeepThings~\cite{zhao2018deepthings} and DeeperThings~\cite{stahl2021deeperthings}, though they fuse layers to reduce transmission delay, they only consider equal-split of fused layers (layer-volume) by assuming that the devices are homogeneous. For AOFL~\cite{zhou2019adaptive}, it split layer-volumes based on a linear ratio. However, as shown in Fig.~\ref{fig:nonlinear}, when the relationship between computing latency and layer configuration is nonlinear, splitting layer-volumes with a linear ratio cannot balance the computing latency. Instead, it can cause computing delay. For \textit{offload}, as it only utilizes one device to compute the CNN model, the end-to-end latency can be larger than the distribution methods that compute the layers in a parallel manner on multiple devices.



We demonstrate the maximum transmission latency and the maximum computing latency among the four devices of Group-DB (50Mbps WiFi) with different distribution methods. As discussed above, distributions with CoEdge~\cite{zeng2020coedge}, MoDNN~\cite{mao2017modnn}, and MeDNN~\cite{mao2017mednn} have large transmission latency due to layer-to-layer data transmission. DeepThings~\cite{zhao2018deepthings} and DeeperThings~\cite{stahl2021deeperthings} have large computing latency because they equally split layer-volumes among devices. As the computing capability of Nano is much smaller than that of Xavier, equally splitting causes high computing delay. AOFL~\cite{zhou2019adaptive} splits layer-volumes with a linear ratio, which can also cause computing delays on devices. In comparison, DistrEdge implicitly learns the nonlinear relationship between the computing latency and layer configurations with DRL. It makes split decisions based on the observations of intermediate latency of previous layer-volumes and minimizes the expected end-to-end latency.

\begin{figure}[!ht]
  \centering
  \vspace{-0.1in}
  \includegraphics[width=0.5\linewidth,height=1in]{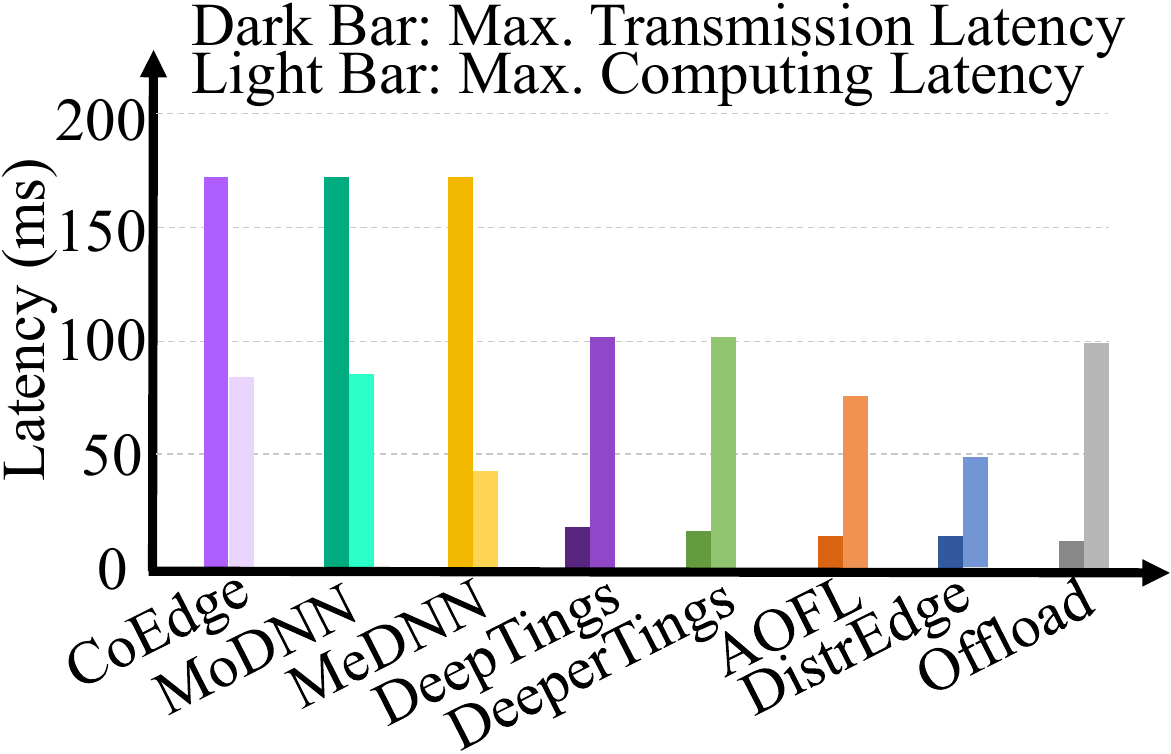}
  \vspace{-0.1in}
  \caption{\small \textit{Max. Transmission Latency (ms) and Max. Computing Latency (ms) among Four Devices (DB, 50Mbps) w/ Different Methods.}}\label{fig:trans_comp_db}
  \vspace{-0.1in}
\end{figure}

\section{Discussion}
\noindent We address the following points in the discussion: 

\noindent(1) State-of-the-art methods (CoEdge and AOFL) that are applicable highly dynamic network environment makes online split decisions based on the monitored network throughput. Similarly, DistrEdge can also make online split decisions (as discussed in Section~\ref{sec:dynamic}) by keeping the actor network online. When network condition changes significantly, AOFL and DistrEdge all need to update their distribution strategy (including partition locations). Benefiting from the lightweight LC-PSS, DistrEdge can update its distribution strategy faster than AOFL based on our evaluation (Section~\ref{sec:dynamic}). 

\noindent(2) Based on the network and edge devices' conditions, DistrEdge divides a CNN model into several split parts. Then it allocates these parts to the edge devices and establishes necessary connections among them. When a user requests CNN inference services, DistrEdge tells the user how to split their input images and which edge devices to send the split images. After the corresponding edge devices generate results, they send the results back to the user. Note that it is possible to obtain an empty set for an edge device. For example, the Pi3 in Group-DC (Table~\ref{tab:heter_devices}) is not allocated with any computation due to its low computing capability.

\noindent(3) TensorRT is one of the popular deep learning acceleration frameworks provided by NVIDIA. We utilize it in our tests to demonstrate that DistrEdge is compatible with other deep learning acceleration techniques. In general, DistrEdge can be utilized alone without any acceleration technique or can be utilized together which other acceleration techniques. 

\noindent(4) We do not consider memory constraint in DistrEdge because state-of-the-art edge devices can provide sufficient memory space for most state-of-the-art CNN inferences~\cite{bianco2018benchmark}. Specifically, the state-of-the-art CNN models consume less than $1.5$GB of memory. In contrast, the state-of-the-art edge devices are equipped with the memory size of $4$GB to $32$GB. In other words, even running a whole CNN model on one edge device does not suffer from memory limitation.

\noindent(5) In this paper, we only study DistrEdge with one dimension split. In the future, we will explore multi-dimension split (e.g., width and height).

\section{Conclusion}
In this paper, we proposed DistrEdge, an efficient CNN inference method on distributed edge devices. DistrEdge consists of a CNN partitioner (LC-PSS) and an LV splitter (OSDS) to find CNN inference distribution strategy. We evaluated DistrEdge in laboratory environments with four types of embedded AI computing devices, including NVIDIA Jetson products. Based on our evaluations, we showed that DistrEdge is adaptive to various cases with heterogeneous edge devices, different network conditions, small/large-scale edge devices, and different CNN models. Overall, we observed that DistrEdge achieves 1.1 to 3$\times$ speedup compared to the best-performance baseline method.



\printbibliography

\end{document}